# Enhanced Curie temperature and room-temperature 50-nm skyrmions achieved in hexagonal ferromagnet $Mn_5Ge_{3+x}$ synthesized via a high-pressure method

Yongsen Zhang, Wei Liu, Meng Shi, Shuisen Zhang, Sheng Qiu, Yaodong Wu*, Jialiang Jiang, Huanhuan Zhang, Hui Han, Kang Wang*, Dingfu Shao, Zhenfa Zi, Chao Ma, Haifeng Du, Mingliang Tian, Shouguo Wang*, and Jin Tang*

Y. Zhang, S. Wang

Anhui Provincial Key Laboratory of Magnetic Functional Materials and Devices, School of Materials Science and Engineering, Anhui University, Hefei 230601, China

Email: sgwang@ahu.edu.cn

W. Liu, H. Han

Institutes of Physical Science and Information Technology, Anhui University, Hefei 230601, China

M. Shi, S. Zhang, S. Qiu, K. Wang, H. Du, M. Tian

Anhui Province Key Laboratory of Low-Energy Quantum Materials and Devices, High Magnetic Field Laboratory, HFIPS, Chinese Academy of Sciences, Hefei, Anhui 230031, China

Email: wangkang@hmfl.ac.cn

Y. Wu, Z. Zi

School of Physics and Materials Engineering, Hefei Normal University, Hefei, 230601, China

Email: wuyaodong@hfnu.edu.cn

J. Jiang, H. Zhang, M. Tian, J. Tang

School of Physics and Optoelectronic Engineering, Anhui University, Hefei, 230601, China

Email: jintang@ahu.edu.cn

D. Shao

Key Laboratory of Materials Physics, Institute of Solid State Physics, HFIPS, Chinese Academy of Sciences, Hefei 230031, China

C. Ma

College of Materials Science and Engineering, Hunan University, Changsha, 410082, China

## Abstract

The development of new high-temperature ultrasmall-size skyrmion materials holds immense significance for the promising applications of topological spintronic devices. In this study, we demonstrate that a high-pressure synthesis technique can significantly elevate the Curie temperature of $Mn_5Ge_{3+x}$ crystals, from 294 K to 350 K. This enhancement is attributed to the combined effects of lattice contraction and increased Ge content, the conclusion supported by Density Functional Theory calculations. Additionally, our real-space magnetic imaging reveals the stability of dipolar skyrmions with diameters of approximately 50 nm at room temperature. Our micromagnetic simulations closely replicate the diverse experimental topological magnetic textures observed. Furthermore, magnetotransport measurements indicate the potential for the electrical distinction between various topological magnetic textures in skyrmion-based devices. We also report deterministic manipulations on single dipolar skyrmions in confined nanostructures by using in-plane currents. The observation, electrical manipulation, and electrical detection of room-temperature ultrasmall topological magnetic textures underscore the potential of $Mn_5Ge_{3+x}$ as a promising platform for spintronic device applications.

# 1. Introduction

Magnetic skyrmions, as topologically protected spin textures, have garnered significant interest for their potential applications in future spintronic devices[1-5]. These nanoscale spin configurations, resembling vortex-like structures, hold the promise of ultra-low power consumption and high-density data storage due to their resilience against external perturbations and their facile manipulation with electric currents[6-9]. The stability of skyrmions at room temperature is crucial for their practical implementation in technological devices[10], [11]. A reduction in the size of room-temperature skyrmions means that more data can be stored within the same area, which is particularly important for high-density storage media and the miniaturization of devices[2], [3] [12-14].

Magnetic skyrmions are commonly observed in materials like FeGe[12] and MnSi[13], which feature a non-centrosymmetric B20-type cubic crystal structure. This distinctive structure promotes the Dzyaloshinskii-Moriya interaction (DMI), thus facilitating skyrmion stabilization. Magnetic skyrmions have been found to have the smallest size down to sub-10 nm in non-centrosymmetric magnets such as $GaV_4S$[15], centrosymmetric magnets such as $Gd_2PdSi$[16], and magnetic multilayers such as Fe/Ir(111)[17], but they stabilize only at very low temperatures. Recent studies have explored room-temperature skyrmion materials, including non-centrosymmetric magnets like CoZnMn[18], MnPtPdSn[19], and FeNiPdP[20], magnetic multilayers[21], and centrosymmetric uniaxial magnets[22-24]. However, the size of room-temperature chiral skyrmions is generally above 100 nm, with only a few reports of sizes below 50 nm in Pt/Co/Ir multilayer films[21]. Therefore, the exploration and identification of new room-temperature 50-nm skyrmion materials remain a critical research objective.

In this study, we report a new room-temperature 50-nm-size skyrmion-hosting hexagonal ferromagnet, $Mn_5Ge_{3+x}$. Combining Density Functional Theory (DFT) calculations, we propose that lattice contraction and increased Ge content induced by the high-pressure (HP) growth method could elevate the Curie temperature of $Mn_5Ge_{3+x}$ from near-room temperature to above room temperature (350 K). Through Lorentz transmission electron microscopy (Lorentz-TEM), we observed room-temperature nanoscale dipolar skyrmions. Additionally, through magnetotransport measurements, we identify the potential for magnetoresistance detection of magnetic textures. Finally, we demonstrate the deterministic manipulation of individual dipolar skyrmions in confined nanostructures using in-plane currents. These findings elucidate the formation mechanisms of ultrasmall room-temperature topological magnetic structures within ferromagnetic $Mn_5Ge_{3+x}$, paving the way for the development of novel spintronic devices.

## 2. Results and Discussion

### 2.1 Enhancement of Curie temperature in high-pressure synthesized $Mn_5Ge_{3+x}$

MnGe has recently been confirmed as a material capable of hosting skyrmions at low temperatures, existing in a non-centrosymmetric phase wherein one of the smallest known skyrmions is formed[25-27]. Additionally, Mn-Ge phases encompass some centrosymmetric phases that exhibit magnetic ordering temperatures close to room temperature. For example, $Mn_5Ge_3$[28], [29], a material with a centrosymmetric crystal structure, features gapped Dirac magnons and is renowned for its unique properties as a topological magnon insulator[30]. $Mn_5Ge_3$ also displays uniaxial magnetic anisotropy[31], [32], and the interplay between uniaxial magnetic anisotropy and magnetic dipole-dipole interactions is expected to generate a variety of topological spin configurations[22-24]; however, there are currently few reports in this area. In the manganese germanium series, whether in the non-centrosymmetric phase or the centrosymmetric phases, the Curie temperature is below room temperature. Therefore, there is an urgent need to develop materials capable of hosting small-sized room-temperature skyrmions.

$Mn_5Ge_3$ is a ferromagnetic material that typically exhibits a hexagonal $D8_8$-type crystal structure with a space group of P6/mcm (No.193)[33], [34], and the easy axis aligned along the [001] axis. As shown in **Figure 1**a, the magnetic atoms Mn occupy two distinct crystallographic positions, 4d and 6g, with the magnetic moments of Mn1 (0.250, 0, 0.250) and Mn2 (0.333, 0.667, 0.250) being 1.96 $\mu_B$ and 3.23 $\mu_B$[35], [36], respectively, aligned along the *c*-axis. The Ge atoms also occupy the 6g positions but with different coordinates (0.610, 0, 0).

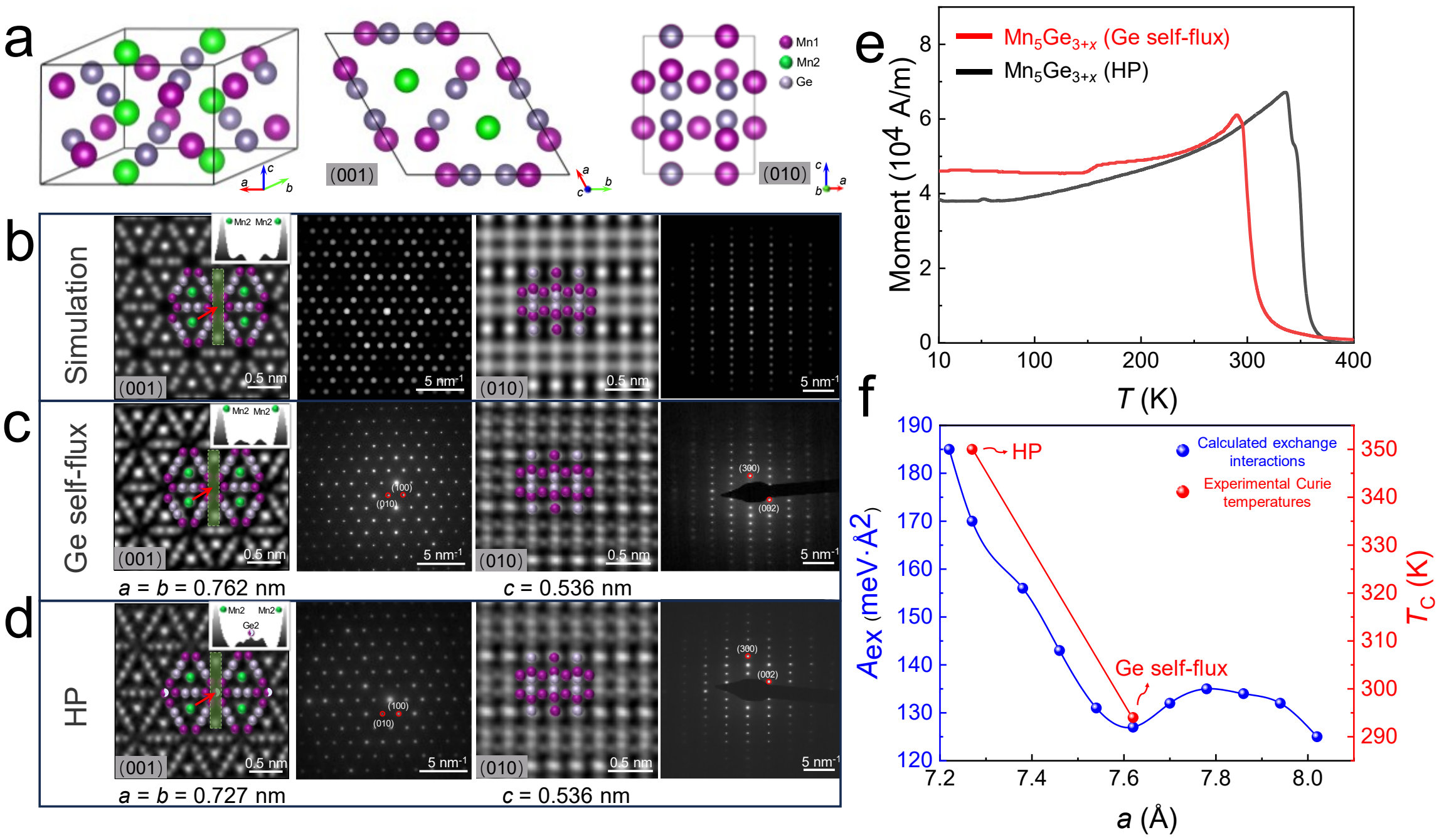

**Figure 1.** a) The crystal structure of $Mn_5Ge_3$, with top-view and cross-sectional views along the *c*-axis and *b*-axis. b) - d) The high-resolution STEM images of the (001) and (010) planes for simulated $Mn_5Ge_3$, $Mn_5Ge_{3+x}$ grown by the Ge self-flux method, and $Mn_5Ge_{3+x}$ synthesized by the HP method, along with the corresponding selected-area electron diffraction (SAED) patterns. e) *M*-*T* curves of $Mn_5Ge_{3+x}$ (growth by the Ge self-flux method) and $Mn_5Ge_{3+x}$ (Synthesis by the HP method) cooled under a constant magnetic field of 20 mT oriented along the [001] direction. f) Calculated exchange interactions of $Mn_5Ge_3$ as a function of the lattice parameter using DFT calculations, with *c* fixed at 5.36 Å, and experimental Curie temperatures for $Mn_5Ge_{3+x}$ crystals grown by the Ge self-flux and the HP methods.

The crystal structure of $Mn_5Ge_{3+x}$ was analyzed using high-angle annular dark-field scanning transmission electron microscopy (HAADF-STEM). Both [001]- and [010]-oriented lamellae (with a thickness of ≈ 45 nm) were prepared by the lift-out method using a focused ion beam (FIB) dual-beam system. Figure 1d presents an atomic-resolution HAADF-STEM image of $Mn_5Ge_{3+x}$ (HP) oriented along the [001] direction. A series of regularly arranged bright and lighter spots represent the positions of atoms. Variations in brightness reflect their respective average atomic numbers, with brighter spots corresponding to heavier elements (Ge) and lighter contrast regions representing lighter elements (Mn)[37], [38]. This allows us to determine the specific occupancies of Mn and Ge atoms, which is further confirmed by X-ray energy-dispersive spectroscopy elemental mapping (EDS-Mapping) analysis (Figure S1). Selected-area electron diffraction (SAED) confirms that this crystal plane is oriented along the [001] direction. When comparing the atomic arrangement of HP synthesized $Mn_5Ge_{3+x}$ (001) with the schematic representations of simulated $Mn_5Ge_3$ (001) and single-crystal $Mn_5Ge_{3+x}$ (001) grown by the Ge self-flux method, we find that the (000) position of the hexagonal network in the $Mn_5Ge_{3+x}$ lattice (marked by the red arrow) is occupied by excess Ge atoms, a fact also supported by the profile of the green rectangle area. EDS analysis also confirmed that the crystal is enriched with Ge (Figure S1 and S2). As shown in Figure 1b, the HAADF-STEM image showing atoms aligned along the [010] orientation displays a basic layer-like arrangement of Mn-Ge atomic columns, which matches perfectly with the $Mn_5Ge_{3+x}$ (010) crystal structure in Figure 1c and 1d, strongly corroborating the high quality of the crystal. Notably, the $Mn_5Ge_{3+x}$ crystals synthesized using the HP method can accommodate a certain degree of compositional fluctuations within the solubility range of the solid solution, thereby exhibiting the presence of Ge-rich phases. However, the fundamental crystallographic framework of these crystals remains consistent with that of typical $Mn_5Ge_3$. To accurately describe the crystal structure characteristics obtained under HP conditions, we introduce the notation $Mn_5Ge_{3+x}$ to represent this structure.

Figure 1e shows the typical magnetization-temperature (*M-T*) curves of $Mn_5Ge_{3+x}$ (Ge self-flux) and $Mn_5Ge_{3+x}$ (HP) cooled under a constant magnetic field of 20 mT oriented along the [001] direction [field-cooling (FC)]. The figure indicates that the $T_C$ of $Mn_5Ge_{3+x}$ (Ge self-flux) is approximately 294 K, near room temperature; whereas the $T_C$ of $Mn_5Ge_{3+x}$ (HP) is approximately 350 K, significantly exceeding room temperature. We speculate that this can be directly attributed to the changes in lattice parameters during the HP synthesis of $Mn_5Ge_{3+x}$[39], resulting in a lattice constant of ($a = b = 0.727$ nm, $c = 0.536$ nm) for $Mn_5Ge_{3+x}$. Compared to $Mn_5Ge_{3+x}$ grown by the Ge self-flux method ($a = b = 0.762$ nm, $c = 0.536$ nm, see Figure S3 for lattice constant determination), the lattice constants (*a* and *b*) are reduced by approximately 4.6%. This alteration impacts the electronic structure of the material, thereby affecting the magnetic interactions between Mn ions[40], [41]. The modification of these interactions allows the material to maintain its ferromagnetic state at higher temperatures. Additionally, differences in the arrangement or chemical environment of Ge atoms also influence the $T_C$. Our EDS experiments reveal that the Ge content slightly increased by 6.6% (to $Mn_5Ge_{3.2}$) following the HP method. DFT calculations indicate that this slight increase in Ge content results in an enhancement of the exchange stiffness constant. Consequently, the rise in $T_C$ observed in the $Mn_5Ge_{3+x}$ crystal can be attributed to the combined effects of lattice contraction and the increased Ge content (Figure 1f and Figure S4). When the lattice parameter of the crystal decreases compared to $a = 7.62$ Å, the exchange interactions increase. The magnetization curves at temperatures of 2 K and 300 K are provided in Figure S5, and we determined the uniaxial anisotropy constant $K_u$ of $Mn_5Ge_{3+x}$ (HP) to be approximately $1.14\times10^5$ $J/m^3$, and the saturation magnetization $M_s$ to be $7.58\times10^5$ A/m at 300 K. No topological spin structures were observed in the $Mn_5Ge_{3+x}$ lamella grown by the Ge self-flux method at room temperature using Lorentz-TEM (Figure S6). All references to $Mn_5Ge_{3+x}$ hereafter pertain to samples synthesized using HP synthesis.

## 2.2 Observation of room-temperature ultrasmall topological magnetic textures in $Mn_5Ge_{3+x}$ lamella

We performed magnetic domain observations on $Mn_5Ge_{3+x}$ lamella cut along the [001] direction using Lorentz-TEM, which enables visualization of in-plane magnetization ($M_{xy}$). As shown in **Figure 2**a, typical stripe domains appear in a 150-nm-thick $Mn_5Ge_{3+x}$ lamella prepared at zero magnetic field and room temperature, with an average width of ≈ 150 nm, as shown in Figure 2b. These domains are stabilized by the combined effect of perpendicular magnetic anisotropy and dipole-dipole interactions. Upon application of an out-of-plane magnetic field, the stripe domains gradually shrink (Figure 2a, 230 mT) and transform into dipolar skyrmions at high fields (Figure 2a, 368 mT). Due to the centrosymmetry of the hexagonal ferromagnet, there is no chiral DMI, leading to the observation of coexisting skyrmions with opposite winding directions[42] (counterclockwise skyrmions appear as black underfocused Fresnel contrast, while clockwise skyrmions appear as white underfocused Fresnel contrast), as shown in Figure 2a and 2c. It is evident that in this state, stripe domains, fragmented magnetic domains, and dipolar skyrmions coexist. With a further increasing magnetic field, the residual stripe domains first gradually evolve into dumbbell-shaped domains and then completely shrink into dipolar skyrmions (Figure 2a, 402 mT). As the magnetic field increases further, the size of the dipolar skyrmions decreases and they ultimately annihilate in the ferromagnetic state. These observations clearly indicate that dipolar skyrmions can be obtained by applying a magnetic field perpendicular to the lamella. The process of entire field-driven evolution is well reproduced in our simulations, as shown in Figure 2f. Additionally, the corresponding TIE analysis fully corroborates the experimental observations (Figure 2d and 2e). Notably, the field-increase process generates a small number of type-II magnetic bubbles (Figure 2a), which undergo a topological transition to type-I magnetic bubbles, also known as dipolar skyrmions (as indicated by red arrows). The topological transition process is also highly reproduced in our simulations (Figure 2f).

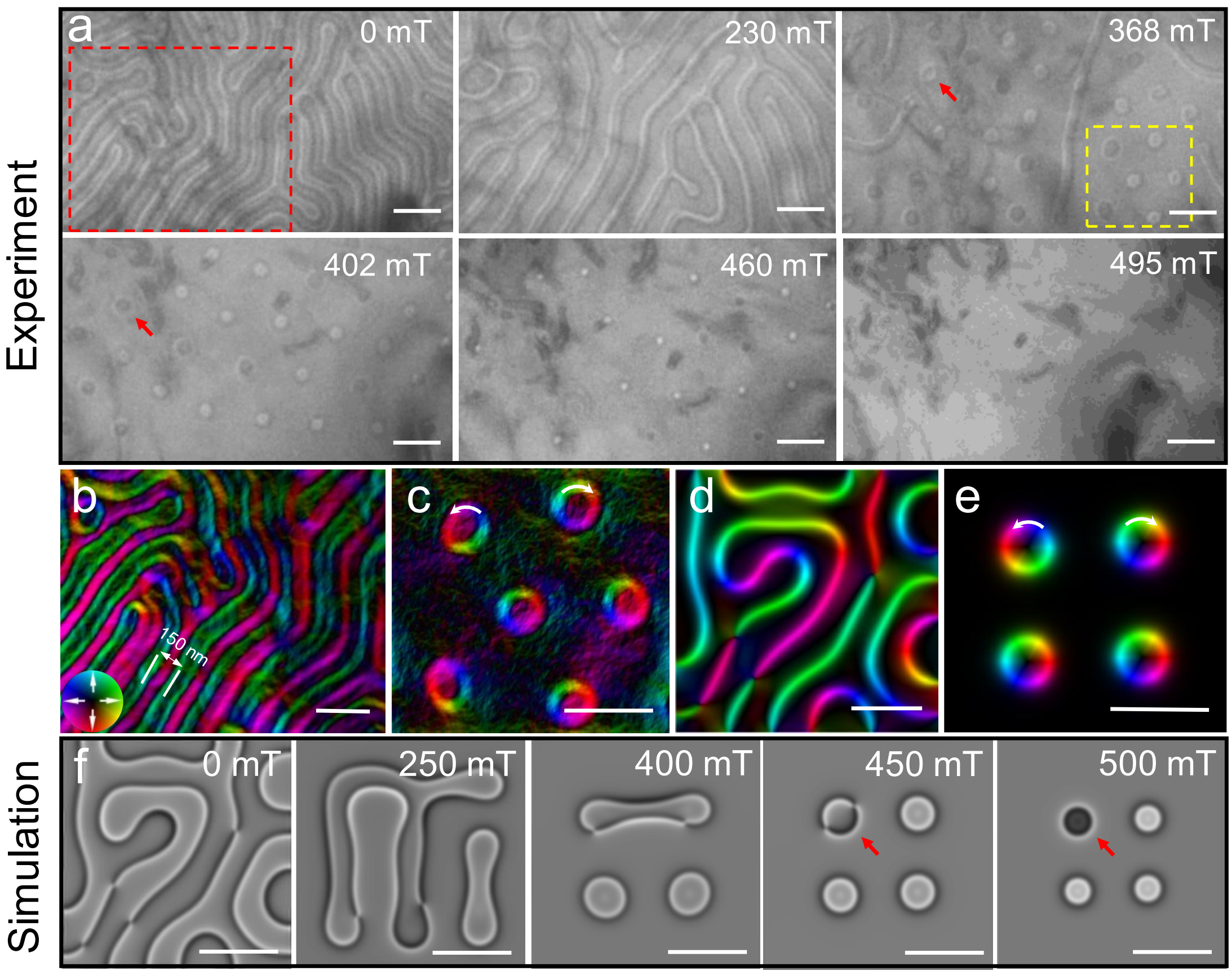

**Figure 2. Observation of dipolar skyrmions above room temperature.** a) Magnetic evolutions from zero-field stripe domains driven by an out-of-plane magnetic field at room temperature. b) and c) Representative in-plane magnetization mappings of stripe domains and dipolar skyrmions retrieved from TIE analysis, correspond to the red and yellow dashed rectangles in a), respectively. d) and e) The in-plane magnetization mappings of stripe domains and dipolar skyrmions, obtained from TIE analysis, as shown in f). f) Corresponding simulated Fresnel images in the field-increasing process. The colorwheel represents the in-plane magnetizations. Defocus, -800 μm. Scale bar, 300 nm.

At room temperature, by first increasing the magnetic field to generate dipolar skyrmions and then reducing the field to zero, the dipolar skyrmions remain stable in the zero magnetic field, as shown in Figure S7. The size of the dipolar skyrmions decreases gradually with increasing magnetic field, ranging from approximately 60 to 150 nm, comparable to those observed in chiral CoZnMn[4] (≈ 115-117 nm), $Fe_{0.5}Co_{0.5}Si$[43] (≈ 90 nm), and in magnetic multilayers[44]. When the applied magnetic field is below 414 mT, the rate of decrease in the size of dipolar skyrmions is relatively small; when it exceeds 414 mT, the rate of decrease markedly increases. As the magnetic field increases to a certain value, the external field gradually dominates, altering the internal energy balance of the skyrmions. To minimize the total system energy, skyrmions adjust their size and shape to adapt to the changes in the external field, which leads to an accelerated reduction in skyrmion size[45], [46]. The dipolar skyrmions ultimately annihilate at 495 mT, as shown in **Figure 3**a. Notably, when the thickness of the lamella is reduced to 50 nm, the period of the stripe domains and the diameter of the dipolar skyrmions during the process of increasing magnetic field remain approximately 50 nm (Figure 3c and Figure S8). During zero-field warming from 100 K to 350 K, the ground-state striped domains maintain strong structural stability, with their period remaining essentially constant at about 150 nm ($t \approx 150$ nm, Figure 3a) and 50 nm ($t \approx 50$ nm, Figure 3c). This indicates that, while thermal fluctuations increase with rising temperature, the contribution of perpendicular magnetic anisotropy is sufficiently large to counteract the thermal fluctuations, thus maintaining the periodicity of the striped domains.

Kittel's Law[47], [48] describes the relationship between the period $d$ of magnetic or polar domains and the film thickness $t$, commonly expressed as $d \propto \sqrt{t}$. The essence of this law lies in how ferromagnetic systems adjust their domain wall configurations to minimize the sum of exchange energy and anisotropy energy, thereby achieving an energetically optimal domain structure. As the sample thickness decreases, the relative contribution of demagnetizing energy to the total energy significantly increases, compelling the system to reduce the stripe domain period to lower the overall energy. Our experimental results are consistent with the trend predicted by Kittel's Law (Figure S9). The dependence of the stripe domain and skyrmion size on the lamella thickness under zero field is summarized in Figure S9, showing an increasing trend with increasing thickness, confirming the dipolar origin as the appearance of skyrmions in $Mn_5Ge_{3+x}$[49-51]. Figure 3e illustrates the generation of approximately 50-nm-sized skyrmions and the field-driven magnetic evolution in a 50-nm-thick lamella. Figure 3b and 3d show the magnetic phase diagrams of dipolar skyrmions as a function of temperature and magnetic field during the field-increasing process in 150-nm and 50-nm thick lamella, respectively. Dipolar skyrmions in $Mn_5Ge_{3+x}$ magnet can survive in a broad field temperature region under field-increasing.

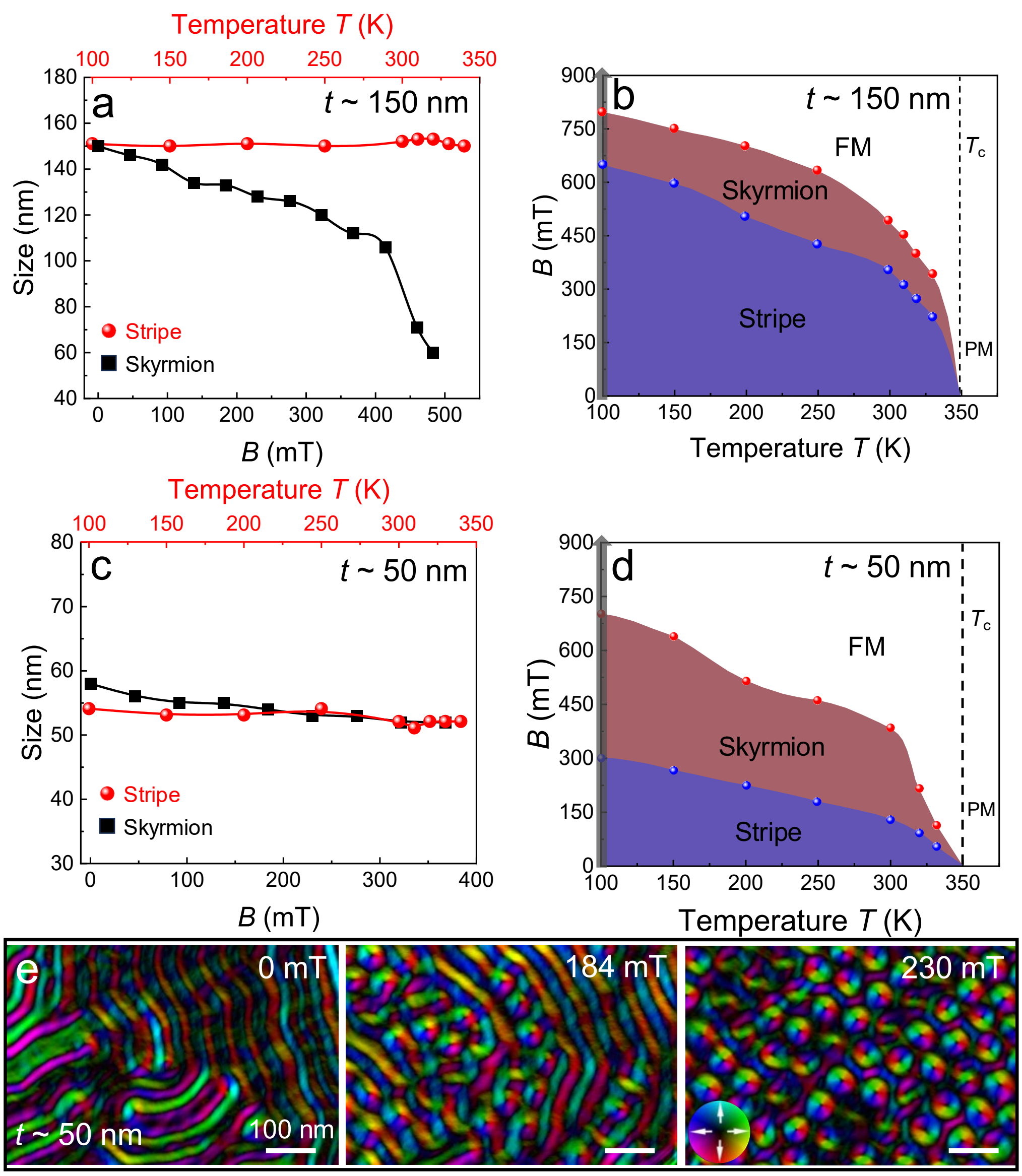

**Figure 3.** Diameter of the dipolar skyrmion as a function of the field at room temperature (black line) and stripe period as a function of temperature $T$ during zero-field warming (red line) in 150-nm a) and 50-nm c) thick $Mn_5Ge_{3+x}$ lamella. b) and d) Magnetic phase diagrams of 150-nm and 50-nm thick $Mn_5Ge_{3+x}$ plate as a function of temperature and magnetic field. e) Field-driven magnetic evolutions in 50-nm thick $Mn_5Ge_{3+x}$ lamella. The colorwheel represents the in-plane magnetizations.

We next investigate the magnetic evolution of skyrmions under reversed magnetic

fields—a common experimental method for generating skyrmion bundles or bags in skyrmion-host materials[7], [52], [53]. Skyrmions formed under negative magnetic fields remain stable even at zero fields (Figure S10). Within a specific range of positive reversed magnetic fields, the outer stripe domains contract, leading to the formation of skyrmion clusters with $Q = 1$, which are encircled by multiple stripe domains. As the magnetic field strength increases further, the inner skyrmions are annihilated before the development of closure stripe domains. Consequently, achieving complete bag-like configurations (Figure S10c) in $Mn_5Ge_{3+x}$ crystals proves challenging under these conditions.

## 2.3 Electrical detection of topological magnetic textures in $Mn_5Ge_{3+x}$

Generally, when the three-dimensional spatial dimensions of a crystal are reduced to a characteristic length scale within the crystal, the interaction between these dimensions and the crystal's characteristic length can lead to the emergence of new quantum phenomena[54]. We further explored the transport properties of $Mn_5Ge_{3+x}$ crystals at the mesoscopic scale. As shown in **Figure 4**a, using a FIB in conjunction with a custom-designed transfer technique, we fabricated microstructured nanostripe devices on intrinsic silicon substrates equipped with template electrodes (Figure S12). At low temperatures, the $Mn_5Ge_{3+x}$ magnet exhibits metallic conductivity properties, and anomalies were detected in the *R*-*T* (Figure 4a) curve around 350 K, where the temperature of resistance transition aligns well with the $T_C$ point, confirming this point as the magnetic transition from the ferromagnetic to the paramagnetic state.

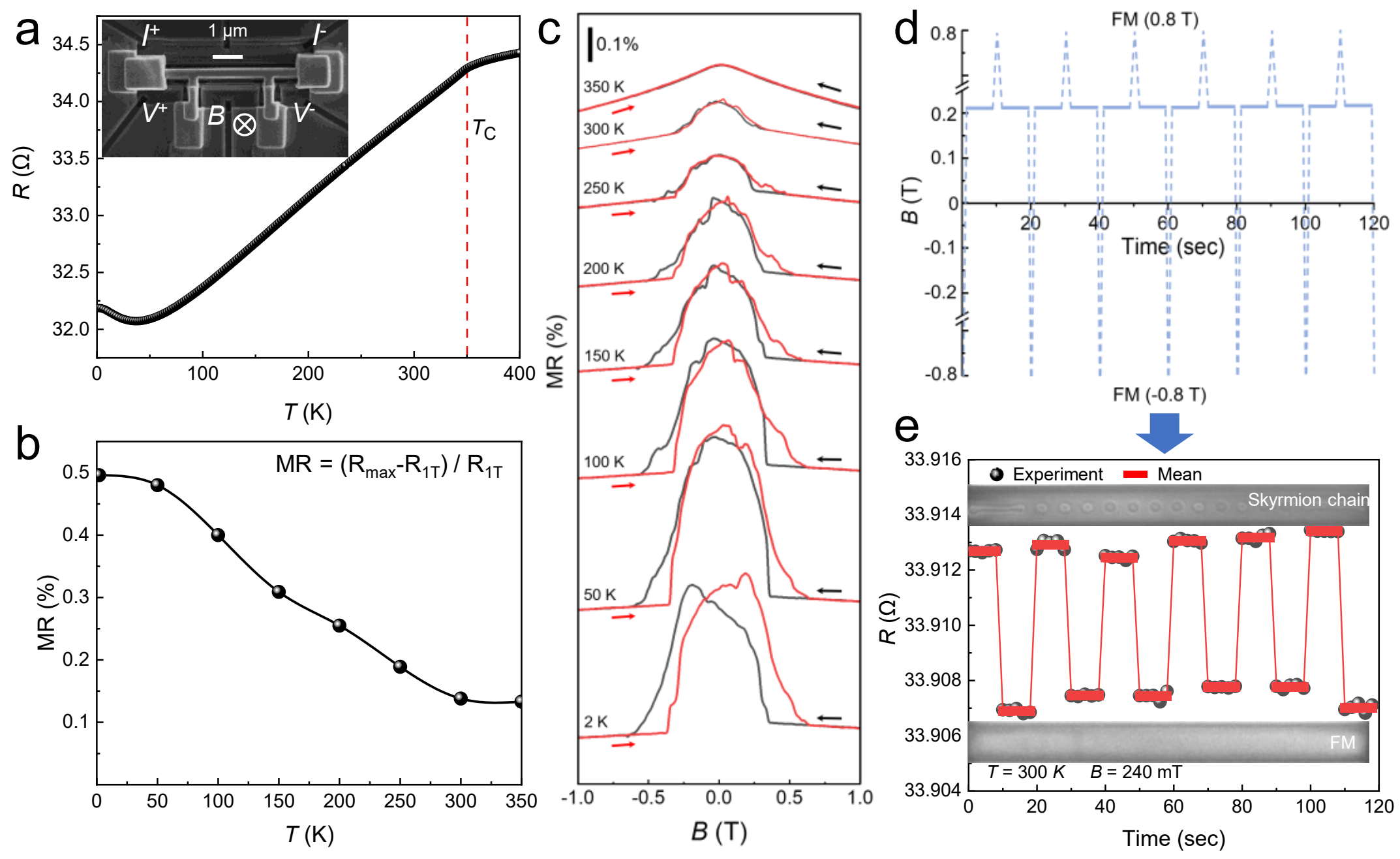

**Figure 4. Magnetotransport properties related to the magnetic texture evolution.** Magnetotransport properties related to the magnetic texture evolution. (a) *R*-*T* curve. The inset is an experimental diagram of a typical four-terminal device. (b) The temperature dependence of the maximum magnetoresistance ($MR_{max}$) at fixed temperatures. (c) MR as a function of the magnetic field at different temperatures. The magnetic field range sweeps from 1 T to -1 T (black line) and back to 1 T (red line). The arrow denotes the direction of the applied magnetic field. (d) The protocol for applying the perpendicular magnetic field as a function of time. (e) The longitudinal resistance-time dependence is acquired according to the magnetic field application method described in (d). The inset in (d) shows the Lorentz Fresnel images of the corresponding magnetic states.

To gain deeper insights into the evolution of magnetic structures at the nanoscale, we conducted magnetoresistance (MR) measurements at various temperatures with the magnetic field oriented such that $B /\!/ c$ (Figure 4c). The MR of ferromagnetic metals typically exhibits a symmetrical relationship with the applied magnetic field, i.e., $R_{xx}$(H) = $R_{xx}$(-H). As shown in Figure 4c, in the temperature range of 100-300 K, the raw MR displays asymmetric peaks. To further investigate this peculiar phenomenon, we applied a vertical magnetic field to the microdevice at 300 K following the method outlined in Figure 4d. Specifically, we initially applied a fixed magnetic field of 0.24 T for approximately 8 seconds, followed by a rapid switch to -0.8 T, inducing a ferromagnetic state. We then quickly returned the field to 0.24 T for another 8 seconds before rapidly increasing it to 0.8 T and subsequently setting it back to 0.24 T, repeating this cycle. The resulting changes in resistance are shown in Figure 4e, where the resistance variations exhibit a largely periodic relationship with the cycling magnetic field, and the corresponding magnetic configurations during this process were characterized using Lorentz-TEM, as shown in Figure 4e. A more detailed correspondence between magnetotransport signals and magnetic structural states is presented in the Figure S13.

The magnetoresistance changes can be attributed to distinct magnetic states—specifically, a single skyrmion chain and a uniform magnetization state—obtained under different field histories, as illustrated in Figure 4e. A direct correlation between the measured transport properties and magnetic domain configurations is further demonstrated in **Figure 5**. Starting from the saturated magnetization state, reversing the magnetic field reveals stable uniform magnetization for $B > 305$ mT (Figure 5c). Horizontal stripe domains then emerge, which subsequently transform into zigzag-like stripe domains at zero fields. Upon further field reversal, a single skyrmion chain stabilizes across the magnetic field interval spanning from −365 mT to −485 mT, progressively transitioning to a ferromagnetic (FM) state at $B < -530$ mT. This magnetic evolution is reproducibly observed during the field-swapping process from −1 to 1 T. The magnetoresistance corresponding to these domain configurations (Figure 5c) is highlighted in Figure 5a. To illustrate the electrical detection of topological domains, two representative states (labeled 2 and 17 in Figure 5b) are analyzed at a fixed field of 380 mT. State 2, representing the FM phase, is achieved by decreasing the field from 1000 to 380 mT. Conversely, State 17, corresponding to the single-skyrmion-chain configuration, is obtained by increasing the field from −1000 m T to 380 mT. The single-skyrmion-chain state exhibits a resistance 0.03 Ω higher than that of the FM state, confirming its distinct transport signature. Therefore, the changes in magnetic structure induced by the processes of increasing and decreasing the magnetic field are responsible for the non-reciprocal symmetry of the MR curve. Within the applied magnetic field range (±1 T), the $MR_{max}$ generally exhibits a decreasing trend with increasing temperature, as shown in Figure 4b. The hysteresis of magnetoresistance also becomes more prominent at low temperatures (Figure 4c), suggesting a larger detection signal of topological domains at low temperatures. These provide a feasible approach for the subsequent electrical probing of topological magnetic structures in $Mn_5Ge_{3+x}$.

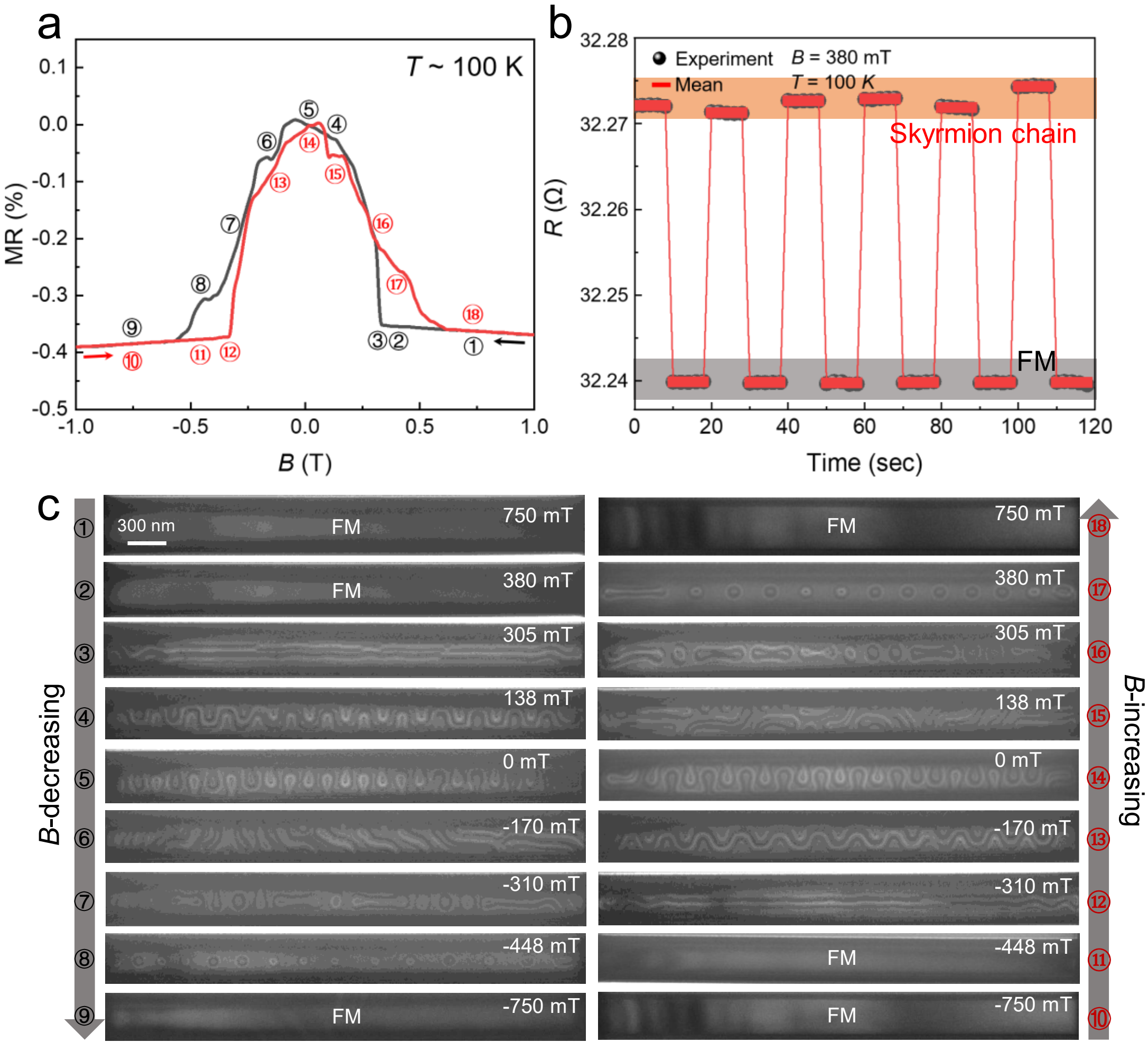

**Figure 5.** (a) MR as a function of the magnetic field at 100 K. The magnetic field range sweeps from 1 T to −1 T (black line) and back to 1 T (red line). The arrow denotes the direction of the applied magnetic field. (b) The longitudinal resistance-time dependence acquired according to the magnetic field application method described in Fig. 4(d). (c) Lorentz Fresnel images depicting the domain structure states under the same conditions as those in panel (a), with the numbers indicating the correspondence.

## 2.4 Electrical manipulation of a single dipolar skyrmion in $Mn_5Ge_{3+x}$ nanostructured cells

For skyrmion-based devices, achieving deterministic control over skyrmions is a critical factor for their practical application[55], [56]. A major challenge in the field of topological spintronics is the precise electrical manipulation of individual topological magnetic structures within strongly confined functional units at room temperature. In the previous section's magnetoresistance-field curves, we clearly observed hysteresis behavior, providing the prerequisite for creating and deleting skyrmions at a fixed field using current.

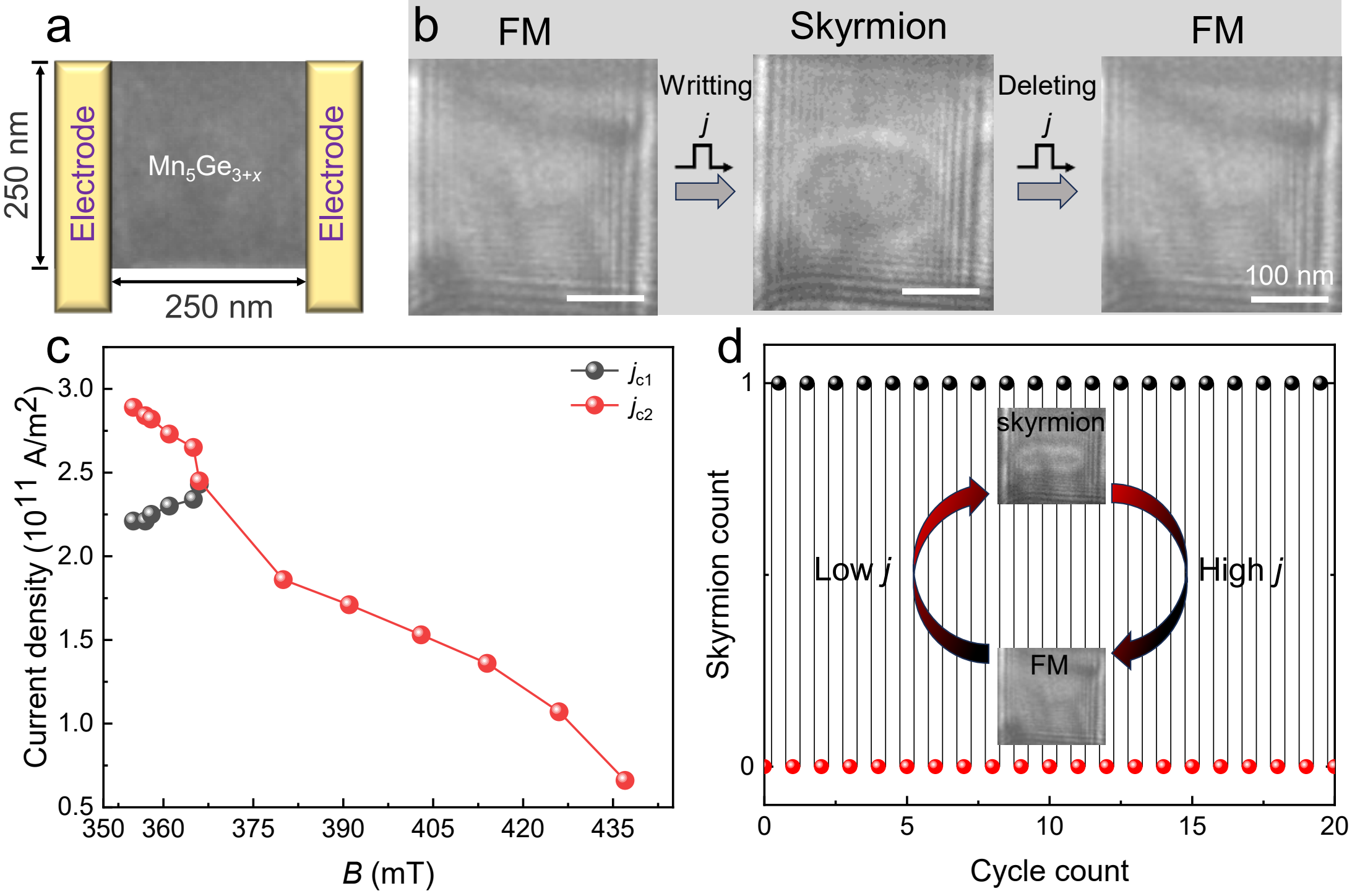

Figure 6. Current-induced creation and deletion of a single skyrmion at room temperature. (a) Schematic diagram of the $Mn_5Ge_{3+x}$ nanostructured cell sample with a (001) plane. (b) Defocused images of different magnetic states in the nanostructured cell after applying pulsed currents. The defocused distance is –300 μm. (c) Dependence of critical current densities $j_{c1}$ and $j_{c2}$ on magnetic field $B$ at a fixed pulse width, $w$ = 5 ns. (d) Cycle count dependence of skyrmion creation and deletion at $B$ = 350 mT. Each cycle count includes a 5-ns low current pulse with a density of ≈ 2.23 × $10^{11}$ A/$m^2$ and a subsequent 5-ns high current pulse with a density of ≈ 2.82 × $10^{11}$ A/$m^2$.

The fabricated 250-nm-wide nanostructured microdevice is shown in **Figure 6**a.

Initially selecting the FM state as the starting point, we explored the evolution of magnetic states under pulsed currents with a pulse width of 5 ns at $B$ = 350 mT (Figure 6b). At low current densities, the sample remains in the FM state until reaching $2.23\times10^{11}$ A/m², after which a skyrmion appears due to the spin-transfer torque (STT) effect induced by the current. With further increases in current density, the skyrmion persists until transitioning back to the FM state at $2.82\times10^{11}$ A/m². This distinct behavior during the transformation of magnetic states under various pulsed currents allows for the unambiguous identification of two critical current densities, $j_{c1}$ and $j_{c2}$. Here, $j_{c1}$ marks the lower critical current density for the transition from the FM state to the skyrmion state, while $j_{c2}$ indicates the higher critical current density for the transition from the skyrmion state back to the FM state. Therefore, the creation and deletion of a single skyrmion can be controlled by applying pulsed currents to the nanostructured cells, encompassing two significant transition processes—from the FM state to the skyrmion state and vice versa.

The deterministic electrical control of single skyrmions is attributed to magnetic domain hysteresis (Figure S14) and combined current-induced effects. In the low-field regions of magnetic domain hysteresis, the skyrmion state represents a more stable phase compared to the uniform magnetization state (Figure S15b). The current-induced spin-transfer torque effect serves as a stimulus to trigger the transition from the metastable FM state to the skyrmion state (Figure S15c). However, when current density further increases, the Joule thermal heating effect becomes dominant. Our results show the thermal cycle from 300 to 320 K and then back to 300 K leads to the skyrmion-to-FM transformation (Figure S15d and Figure S15e).

The critical current densities $j_{c1}$ (for writing) and $j_{c2}$ (for deleting) are highly sensitive to the magnetic field $B$, as shown in Figure 6c. When the magnetic field exceeds 365 mT, $j_{c1}$ may surpass $j_{c2}$, leading to sufficient Joule heating that reduces the saturation magnetization and stabilizes the FM state, making it impossible to transform the FM state into dipolar skyrmions using pulsed currents. On the other hand, with the increasing magnetic field, the stability of the FM state is enhanced for $j_{c2}$, requiring less reduction in saturation magnetization, which results in a decrease in $j_{c2}$, as shown in Figure 6c. Based on the transition process between the FM and skyrmion states, we applied a pulse with a width $w$ of 5 ns current to the microdevice and selected two discrete current densities to establish a write-delete cycle for the nanostructured unit. Under a pulse width of 5 ns, a current density of $2.23 \times 10^{11}$ A/m² was chosen for writing a single skyrmion from the FM state, while a current density of $2.82 \times 10^{11}$ A/m² was selected for reverse conversion. By alternately applying these two pulsed currents to the sample, we achieved deterministic cycles of writing and deleting (Figure 6d; Supplementary Video 1). The successful electrical manipulation of skyrmions at room temperature highlights the potential of $Mn_5Ge_{3+x}$ materials for applications in spintronic devices.

## Conclusion

In conclusion, we successfully elevated the $T_C$ of $Mn_5Ge_{3+x}$ using a HP synthesis method, achieving a ferromagnetic transition above room temperature (≈ 350 K). We obtain stable dipolar skyrmion at room temperature in $Mn_5Ge_{3+x}$ lamella, achieving sizes as small as ≈ 50 nm. Futhermore, we successfully show the deterministic electrical creation and deletion of individual skyrmions in confined $Mn_5Ge_{3+x}$ unit (size ≈ 250 nm × 250 nm) using single nanosecond-scale pulsed currents. Finally, we propose that single skyrmion chains in confined nanostripes can be detected via magnetoresistance measurements. These results not only establish $Mn_5Ge_{3+x}$ as a promising material for topological magnetic studies but also open up new avenues for developing spintronic devices with enhanced functionalities operating at ambient conditions.

## 3. Experimental Section

*Sample preparation*: $Mn_5Ge_{3+x}$ crystals were prepared through two different growth methods. The first method involved the Ge self-flux method. Specifically, Mn granules (99.9% purity) and Ge granules (99.999% purity) were placed in an alumina crucible and then sealed into a vacuum quartz tube. The quartz tube was placed in a furnace and maintained at 1150°C for 1 day to ensure complete dissolution of Mn in the Ge solution. The furnace temperature was then slowly reduced to 850 °C and held at this temperature for 10 hours. Finally, the quartz tube was inverted and inserted into a centrifuge to separate the single crystals from the excess Ge flux.

The second method involved HP synthesis. In this process, slightly more than stoichiometric amounts of high-purity Mn (99.9%) and Ge (99.999%) powders were used. The components were melted in a vacuum arc furnace. The resulting ingot was then annealed under high pressure and high-temperature conditions (8 GPa, 1000°C; model: Max Voggenreiter LPR-1000) for 2.5 hours.

Notably, there exists a degree of flexibility around the stoichiometric ratio of $Mn_5Ge_3$.

The binary phase diagram of the Mn-Ge alloy is provided in the Figure S16. Even when deviating from the stoichiometric ratio, excess Ge atoms can be accommodated within the single-phase matrix through solid solubility, without precipitating secondary phases. This indicates that compositions moderately enriched in Ge can still maintain the desired single-phase structure, as the lattice is capable of accommodating compositional fluctuations within the solubility limits of the solid solution[57].

*Structure characterization*: The HAADF-STEM images and electron diffraction patterns were acquired using a transmission electron microscope (Thermo Scientific Themis Z 3.2) equipped with a spherical aberration corrector and operated at 300 kV.

*Device fabrication and magnetic transport measurement*: First, a slice approximately 12 μm in length, 6 μm in width, and 1 μm in thickness was fabricated on the sample using FIB technology. This slice was then transferred onto an intrinsic silicon substrate with template electrodes using a custom-designed transfer platform. Finally, the targeted size and shape of the six-terminal Hall electrode sample were precisely etched using FIB technology. Magnetic transport measurements were conducted using a Physical Property Measurement System (Quantum Design, PPMS). The magneto-transport properties were measured using a standard six-terminal Hall bar method.

*Preparation of $Mn_5Ge_{3+x}$ lamellas*: The 150 and 50-nm-thick $Mn_5Ge_{3+x}$ nanostructured lamellas were fabricated from a bulk single crystal using a standard lift-out method with an SEM-FIB dual-beam system (Helios Nanolab 600i, FEI).

*Lorentz-TEM measurements*: In-situ Fresnel imaging was performed in a Lorentz-

TEM (Talos F200X, FEI) with an acceleration voltage of 200 kV to investigate magnetic domains in the $Mn_5Ge_{3+x}$ lamellas. The TEM holder (model 636.6, Gatan) supported variable temperature measurements.

*First-principles calculations*: The exchange stiffness describes the energy of a long wavelength spin wave excitation. In particular, if the energy as a function of wave number $q$ is written as $E(q) = E_0 + A_{ex}q^2 + \cdots$, the exchange stiffness constant, $A_{ex}$ is seen to be given by the curvature of the $E(q)$ vs. $q$ curve. $A_{ex}$ can be calculated from the Heisenberg exchange parameters as[58]: $A_{ex} = \frac{2\mu_B}{3M}\sum_{\mathbf{R}}\sum_{ij} J_{i\mathbf{0},j\mathbf{R}} r^2_{i\mathbf{0},j\mathbf{R}}$. Where $r_{i\mathbf{0},j\mathbf{R}}$ is the distance between atom $i$ in cell $\mathbf{0}$ and atom $j$ in cell $\mathbf{R}$.

The magnetic exchange parameters $J_{i\mathbf{0},j\mathbf{R}}$ is calculated based on a real-space Green's function[59-61] implemented in QuantumATK[62], including ***k***-point density of 12 Angstrom along $x$, $y$ and $z$ directions. A default cut-off energy, and ***k***-point density of 7 Angstrom along $x$, $y$, and $z$ directions, Perdew-Burke-Ernzerhof (PBE) functional[63], and the PseudoDojo pseudopotentials[64] are used for the self-consistent calculations. The GGA+U functional is included in the calculations[65].

*Micromagnetic simulation*: The zero-temperature micromagnetic simulations for magnetic domains in 150-nm-thick lamellas were performed using MuMax3. The numerical simulation parameters were specifically tailored to match the material properties of $Mn_5Ge_{3+x}$, including exchange stiffness $A_{ex} = 2.30 \times 10^{-11}$ J m$^{-1}$, saturation magnetization $M_s = 7.58 \times 10^5$ A m$^{-1}$, and uniaxial anisotropy constant $K_u = 1.14 \times 10^5$ J m$^{-3}$. The cell size adopted for the simulations was 4 nm × 4 nm × 4 nm.

## Supporting Information

Supporting Information is available from the Wiley Online Library or from the author.

## Acknowledgements

This work was supported by the National Key R&D Program of China, Grant No. 2022YFA1403603 (H.D.); the Natural Science Foundation of China, Grants No. 52325105 (H.D.), 12422403 (J.T.), U24A6001 (J.T.), 12174396 (J.T.), 12104123 (Y.W.), 12204006 (W.L.), and 12241406 (H.D.); the Anhui Provincial Natural Science Foundation, Grant No. 2308085Y32 (J.T.) and 2408085QA022 (Y.Z.); the Natural Science Project of Colleges and Universities in Anhui Province, Grants No. 2022AH030011 (J.T.) and 2024AH030046 (Y.W.); CAS Project for Young Scientists in Basic Research, Grant No. YSBR-084 (H.D.); Systematic Fundamental Research Program Leveraging Major Scientific and Technological Infrastructure, Chinese Academy of Sciences, Grant No. JZHKYPT-2021-08 (H.D.); Anhui Province Excellent Young Teacher Training Project, Grant No. YQZD2023067 (Y.W.); the 2024 Project of GDRCYY No. 217 (Y.W.); and the China Postdoctoral Science Foundation Grant No. 2023M743543 (Y.W.) and 2024M760006 (Y.Z.).

## Conflict of Interest

The authors declare no conflict of interest.

## Author Contributions

Y.Z. and W.L. contributed equally to this work. S.W., H.D., M.T., and J.T. supervised the project. J. T. conceived the idea and designed the experiments. W.L. and H.H.

synthesized the bulk crystals. Y.Z., H.Z., and K.W. fabricated the microdevices and performed the TEM measurements with the help of J.J. and C.M. Y.Z., S.Q., Y.W., and Z.Z. performed the PPMS measurements. M.S. and S.Z. performed the simulations with the help of D.S. Y.Z., J.T., Y. W., and S.W. wrote the manuscript with input from all authors. All authors discussed the results and contributed to the manuscript.

## Data Availability Statement

The data that support the plots provided in this paper and other finding of this study are available from the corresponding author upon request.

**Supporting Information**

# Enhanced Curie temperature and room-temperature 50-nm skyrmions achieved in hexagonal ferromagnet $Mn_5Ge_{3+x}$ synthesized via a high-pressure method

Yongsen Zhang, Wei Liu, Meng Shi, Shuisen Zhang, Sheng Qiu, Yaodong Wu*, Jialiang Jiang, Huanhuan Zhang, Hui Han, Kang Wang*, Dingfu Shao, Zhenfa Zi, Chao Ma, Haifeng Du, Mingliang Tian, Shouguo Wang*, and Jin Tang*

Y. Zhang, S. Wang

Anhui Provincial Key Laboratory of Magnetic Functional Materials and Devices, School of Materials Science and Engineering, Anhui University, Hefei 230601, China

Email: sgwang@ahu.edu.cn

W. Liu, H. Han

Institutes of Physical Science and Information Technology, Anhui University, Hefei 230601, China

M. Shi, S. Zhang, S. Qiu, K. Wang, H. Du, M. Tian

Anhui Province Key Laboratory of Low-Energy Quantum Materials and Devices, High Magnetic Field Laboratory, HFIPS, Chinese Academy of Sciences, Hefei, Anhui 230031, China

Email: wangkang@hmfl.ac.cn

Y. Wu, Z. Zi

School of Physics and Materials Engineering, Hefei Normal University, Hefei, 230601, China

Email: wuyaodong@hfnu.edu.cn

J. Jiang, H. Zhang, M. Tian, J. Tang

School of Physics and Optoelectronic Engineering, Anhui University, Hefei, 230601, China

Email: jintang@ahu.edu.cn

D. Shao

Key Laboratory of Materials Physics, Institute of Solid State Physics, HFIPS, Chinese Academy of Sciences, Hefei 230031, China

C. Ma

College of Materials Science and Engineering, Hunan University, Changsha, 410082, China

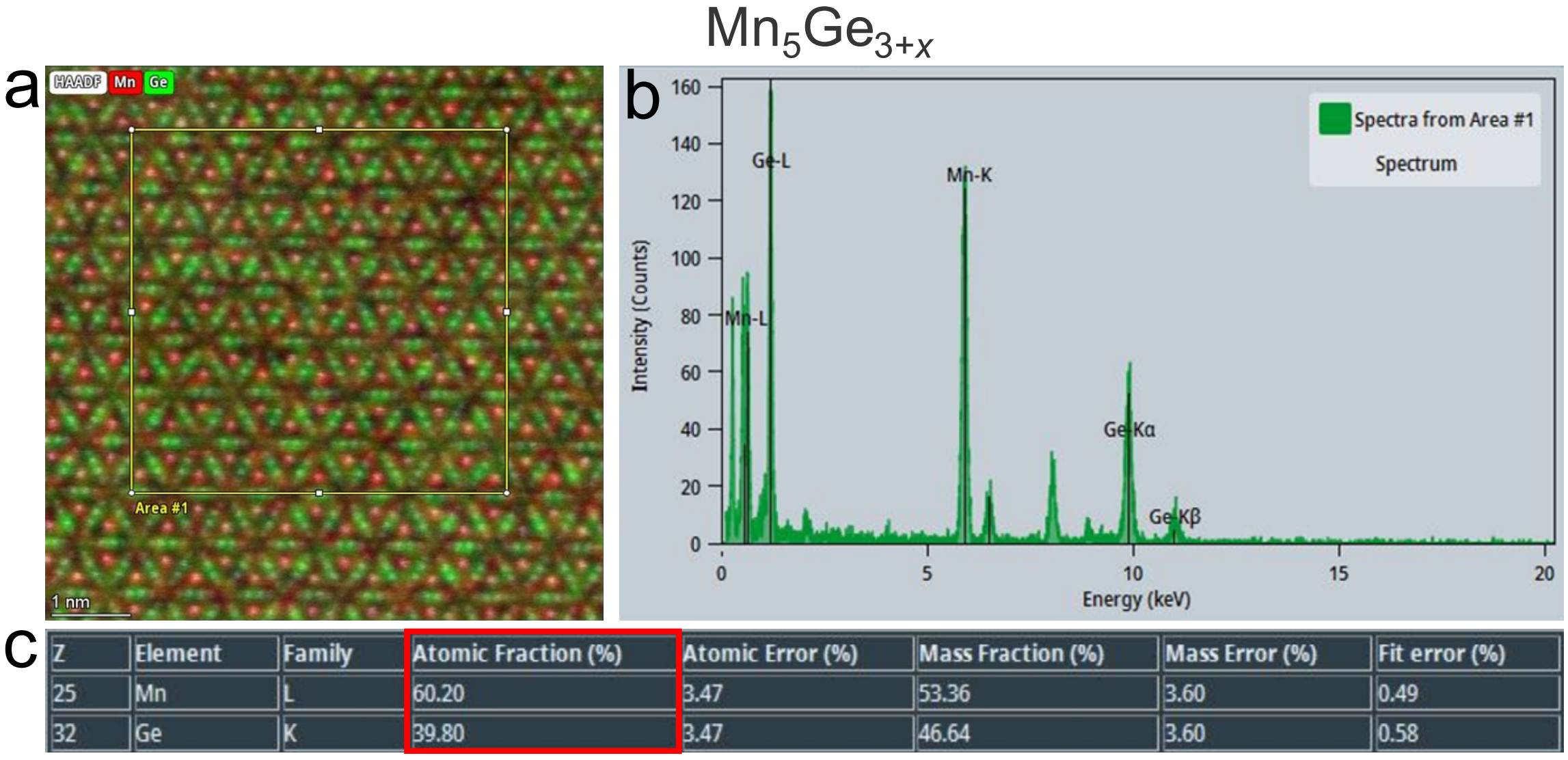

| Z | Element | Family | Atomic Fraction (%) | Atomic Error (%) | Mass Fraction (%) | Mass Error (%) | Fit error (%) |
|---|---|---|---|---|---|---|---|
| 25 | Mn | L | 60.20 | 3.47 | 53.36 | 3.60 | 0.49 |
| 32 | Ge | K | 39.80 | 3.47 | 46.64 | 3.60 | 0.58 |

**Figure S1.** a) The EDS-mapping of $Mn_5Ge_{3+x}$ (HP), illustrating the distribution of Mn and Ge within the sample. b) Spectra from area#1 in a). c) Atomic fraction from area#1 in a).

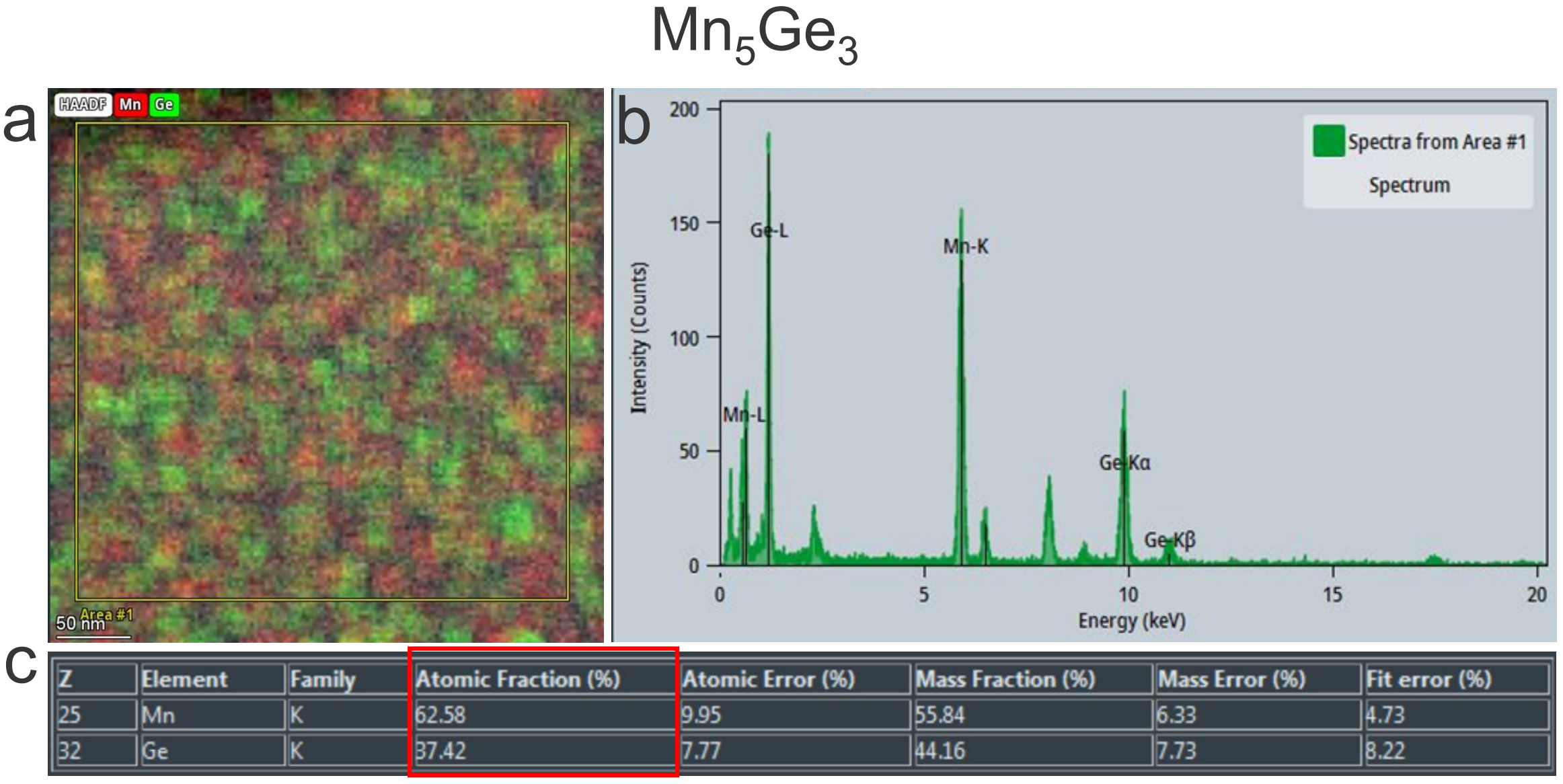

| Z | Element | Family | Atomic Fraction (%) | Atomic Error (%) | Mass Fraction (%) | Mass Error (%) | Fit error (%) |
|---|---|---|---|---|---|---|---|
| 25 | Mn | K | 62.58 | 9.95 | 55.84 | 6.33 | 4.73 |
| 32 | Ge | K | 37.42 | 7.77 | 44.16 | 7.73 | 8.22 |

**Figure S2** a) The EDS-mapping of $Mn_5Ge_3$ (self-flux), illustrating the distribution of Mn and Ge within the sample. b) Spectra from area#1 in a). (c) Atomic fraction from area#1 in a).

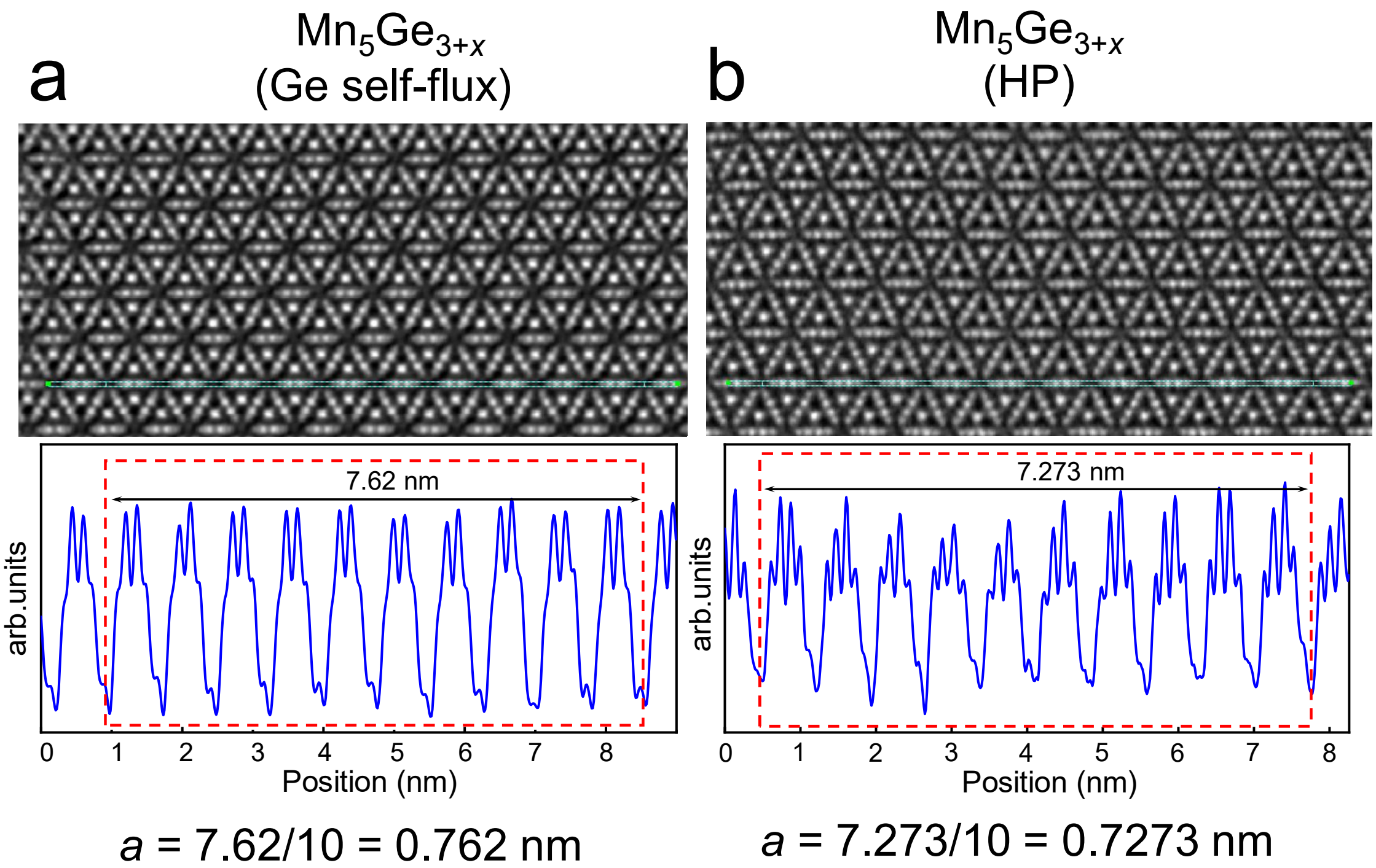

**Figure S3.** To determine the lattice parameters of $Mn_5Ge_{3+x}$ by Ge self-flux (**a**) and $Mn_5Ge_{3+x}$ by HP (**b**), the length of 10 periods was measured and averaged.

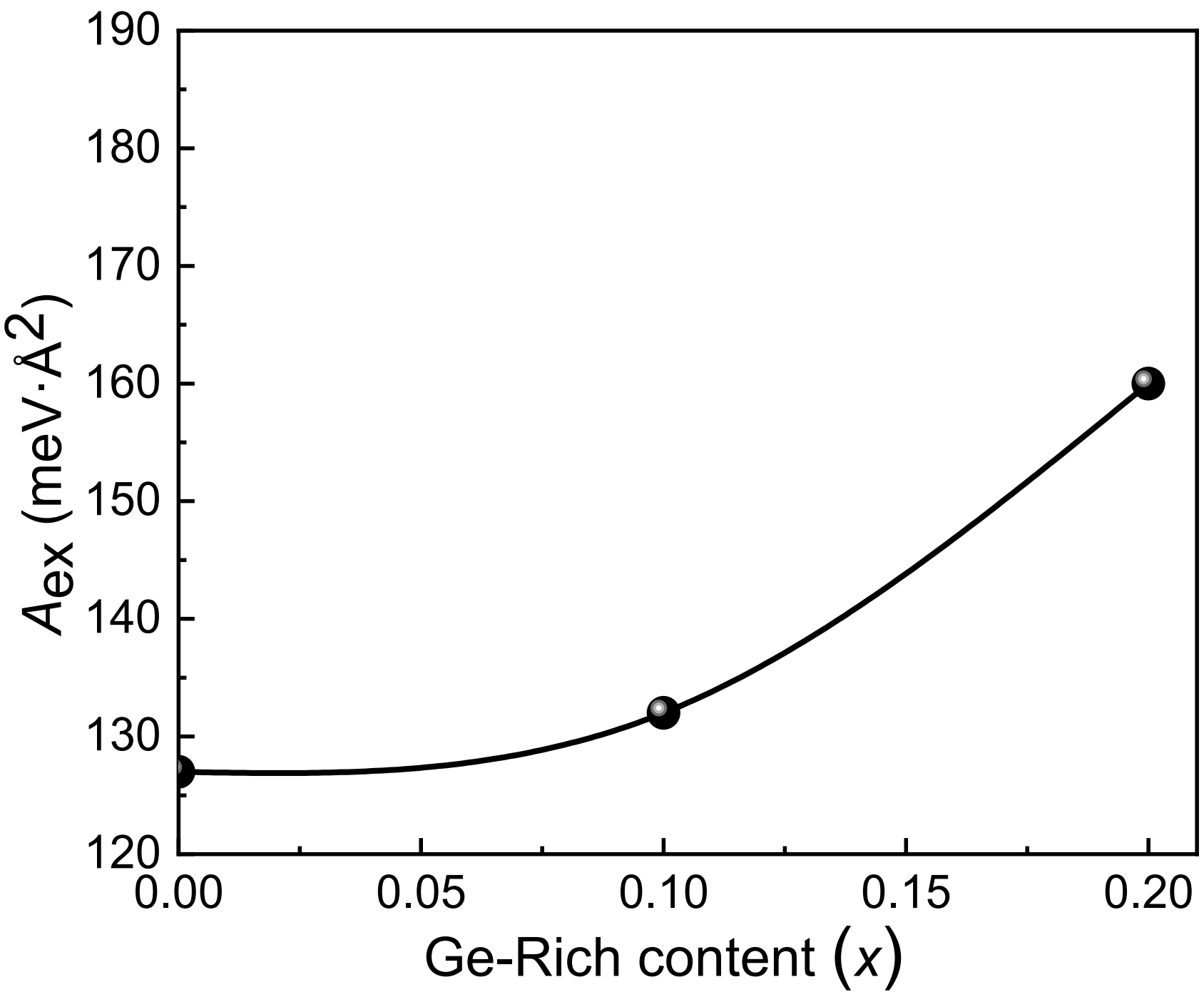

**Figure S4.** Calculated exchange interactions of $Mn_5Ge_3$ as a function of Ge-rich content *x* using DFT calculations.

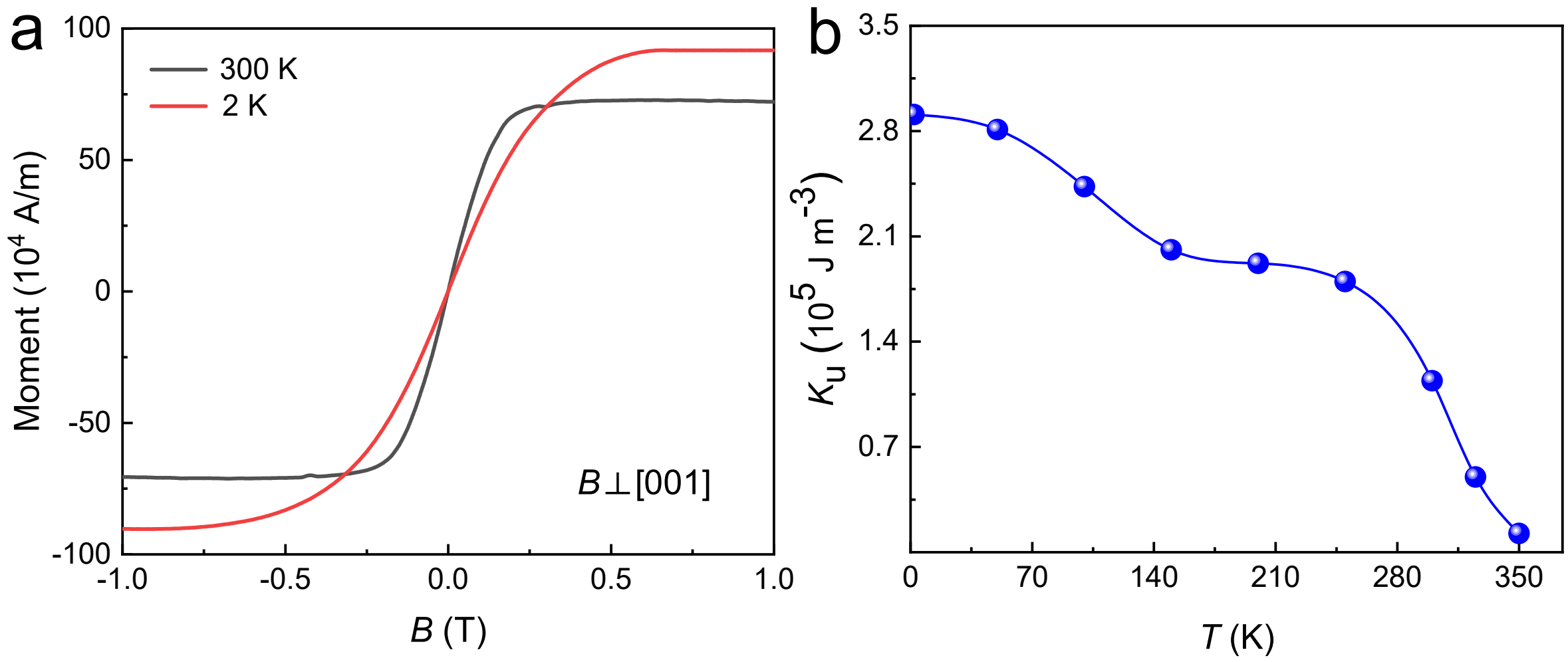

**Figure S5.** a) Magnetic field dependence of magnetization of $Mn_5Ge_{3+x}$ (HP) at 2 K (red line) and 300 K (black line) for $B\perp$[001]. b) The temperature dependence of the magnetic anisotropy constant $K_u$.

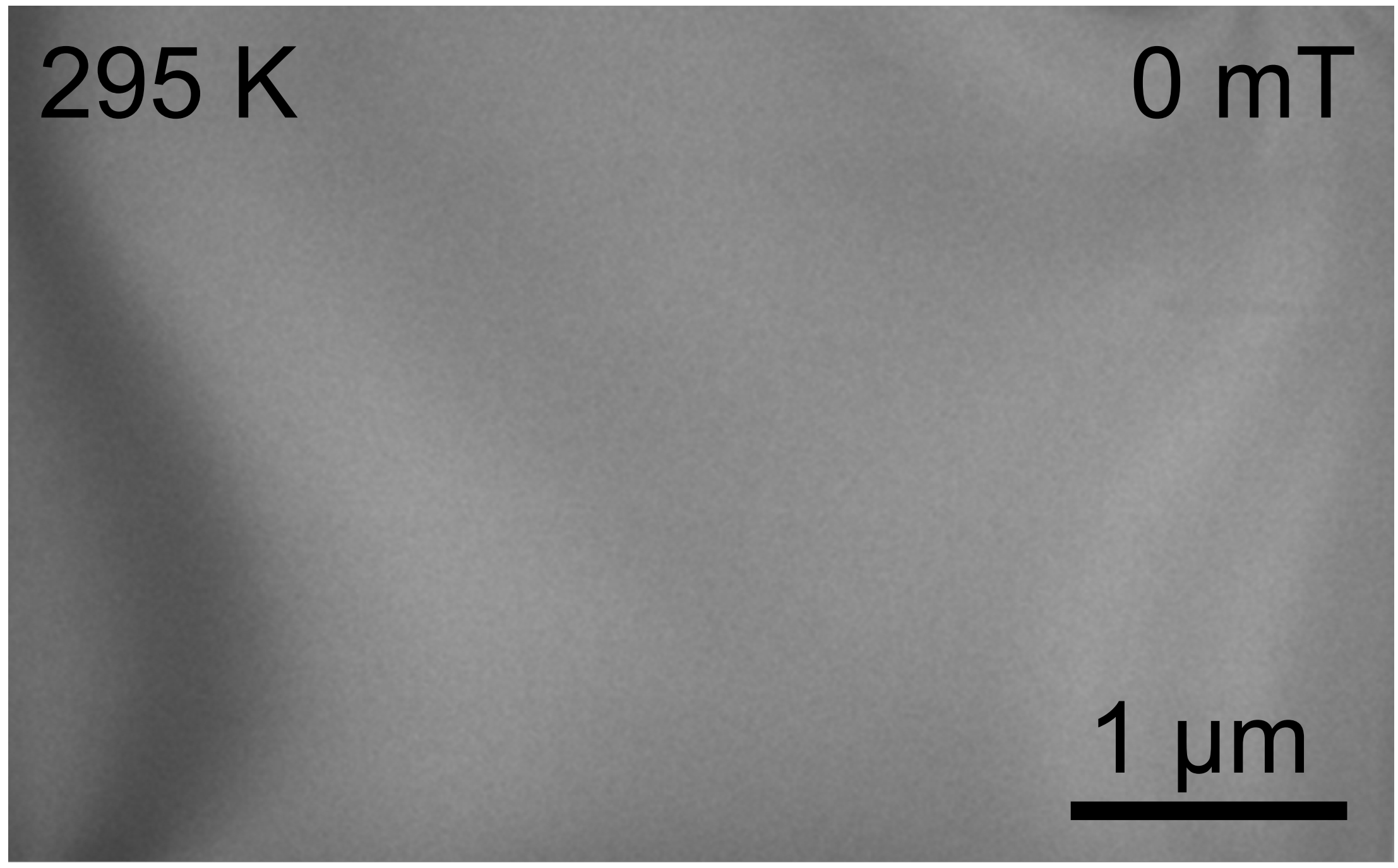

**Figure S6.** Fresnel image of $Mn_5Ge_{3+x}$ sample grown by the Ge self-flux method at room temperature. Defocus, -800 μm.

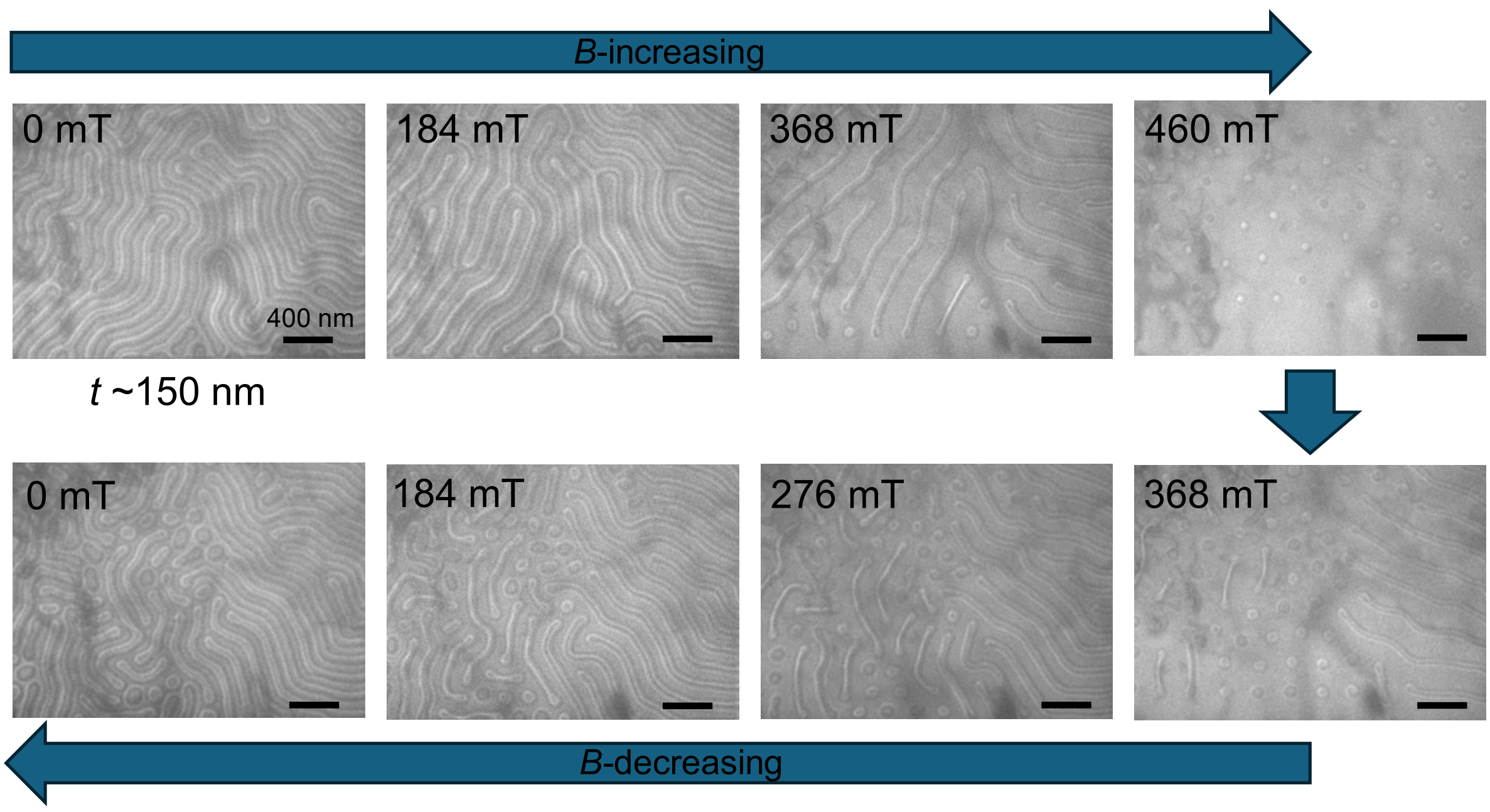

**Figure S7.** The method for stabilizing dipolar skyrmions in zero magnetic field. Defocus, -800 µm.

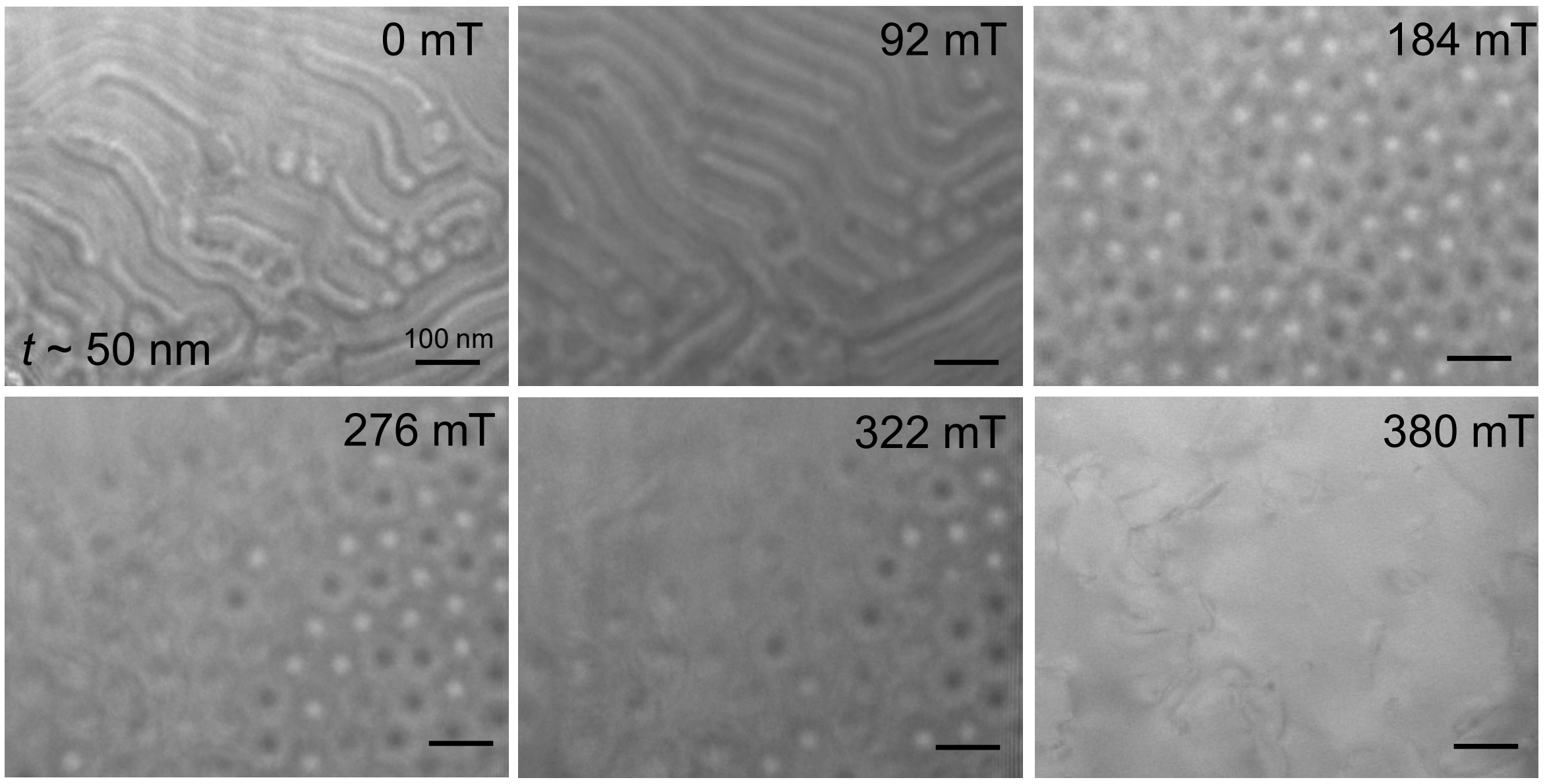

**Figure S8.** Magnetic evolutions from zero-field skyrmions in a 50-nm-thick $Mn_5Ge_{3+x}$ lamella driven by out-of-plane magnetic field at room temperature. Defocus, -800 µm.

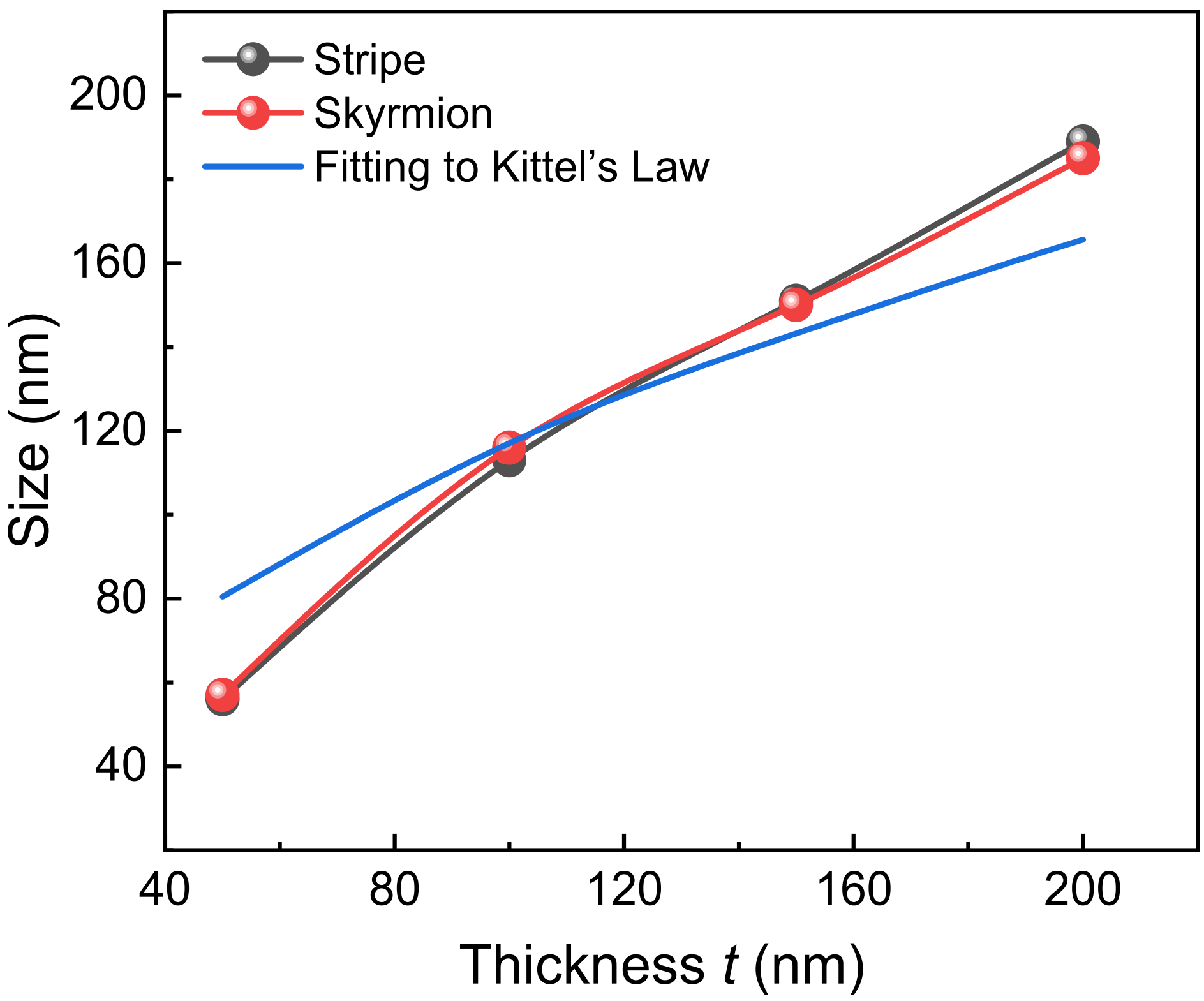

**Figure S9.** The dependence of the stripe domain and skyrmion size on the lamella thickness under zero field at room-temperature.

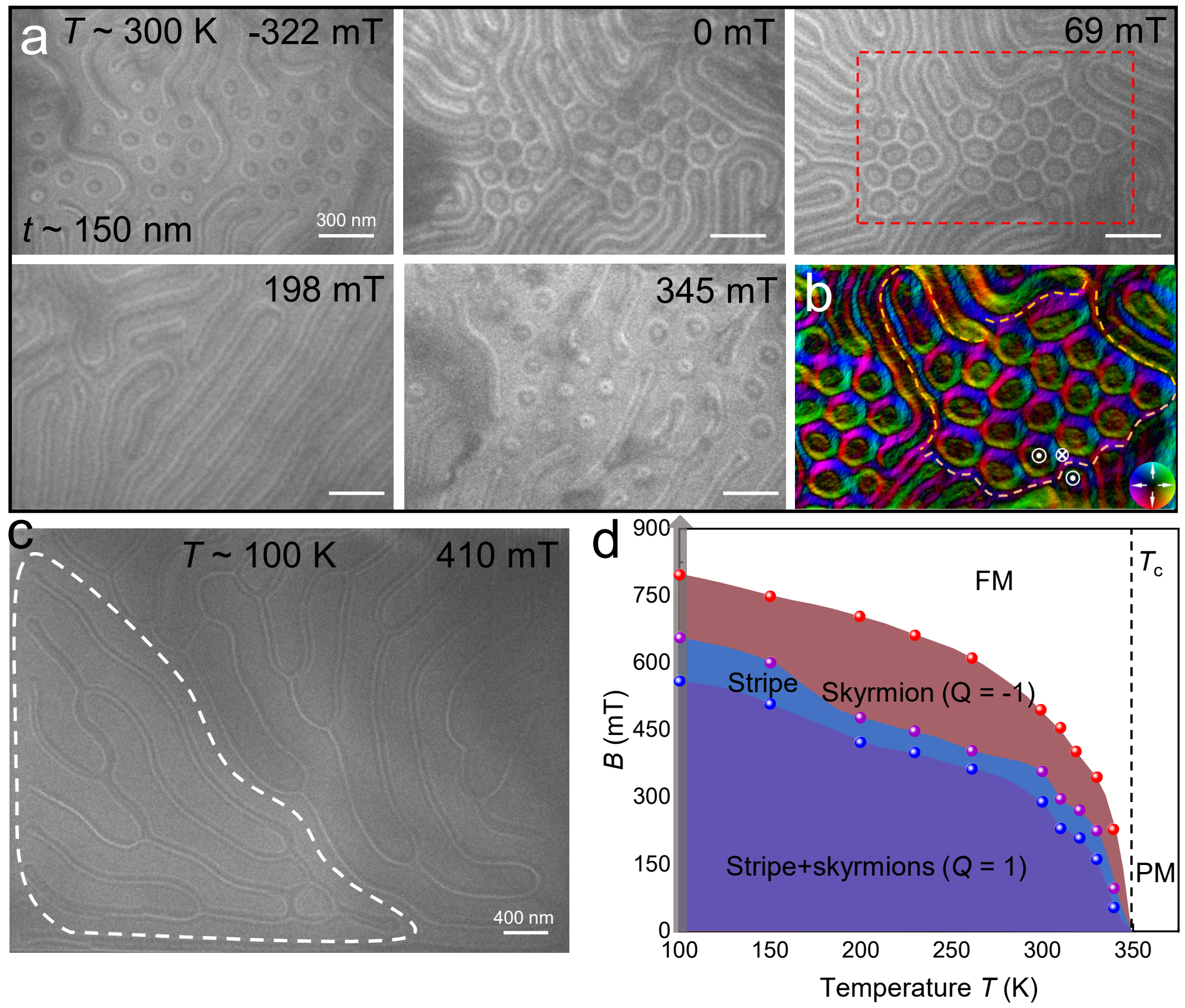

**Figure S10.** Observation of dipolar skyrmions with reversed topological charges enclosed by multiple stripe domains. a) Dipolar skyrmions with reversed topological charges enclosed by multiple stripe domains obtained by reversed magnetic fields. b) Representative in-plane magnetization mappings of dipolar skyrmions with reversed topological charges enclosed by multiple stripe domains retrieved from TIE analysis, corresponds to the red dashed rectangles in a), and the white symbols represent the direction of the magnetic moment. c) The magnetic structure during the reverse magnetic field process at 100 K. The white dashed box indicates the uncontracted bag-like configurations. d) Magnetic phase diagrams of dipolar skyrmions with reversed topological charges enclosed by multiple stripe domains as a function of temperature and magnetic field. The data points in the diagram are the critical points of the two-state transition. FM and PM represent ferromagnetic and paramagnetic states, respectively. The color represents the in-plane magnetization according to the color wheel. Defocus, -800 μm.

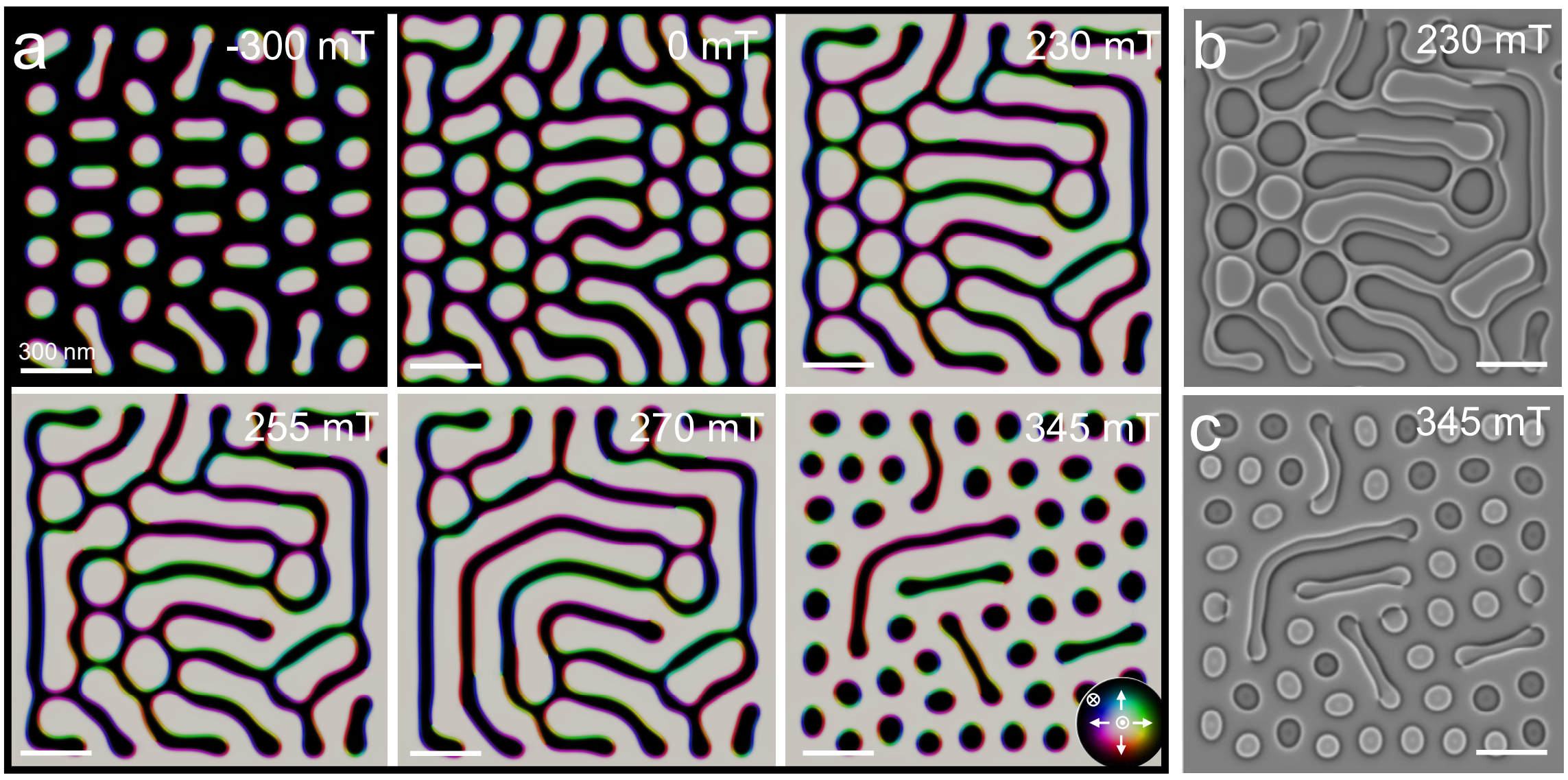

**Figure S11.** a) Simulation of the creation process of dipolar skyrmion bags. b) and c) Simulated Fresnel images corresponding to the magnetic fields of 230 mT and 345 mT in a). The color represents magnetization according to the colorwheel.

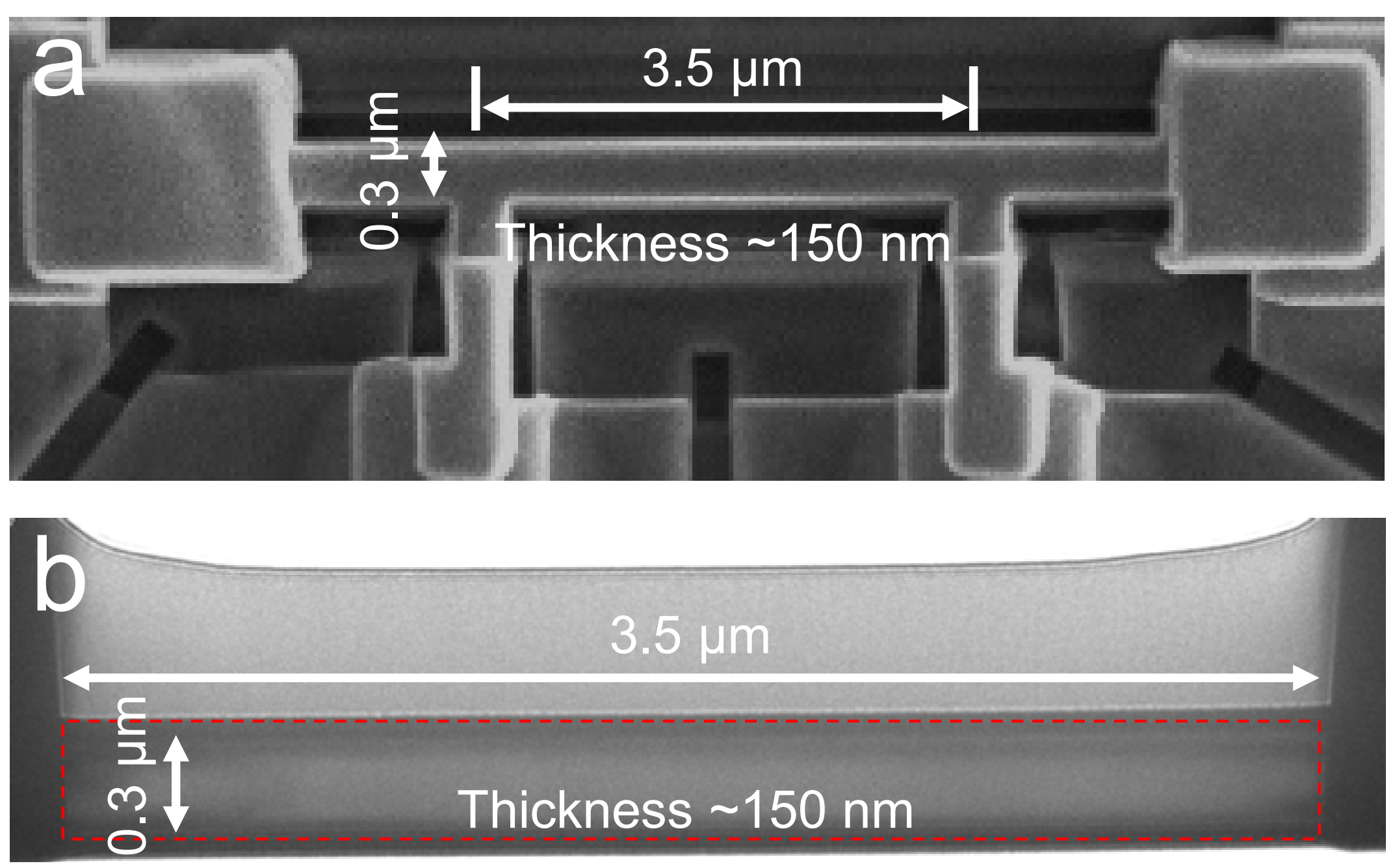

**Figure S12.** a) The microdevice for transport measurements. b) The lamella for observation with Lorentz-TEM.

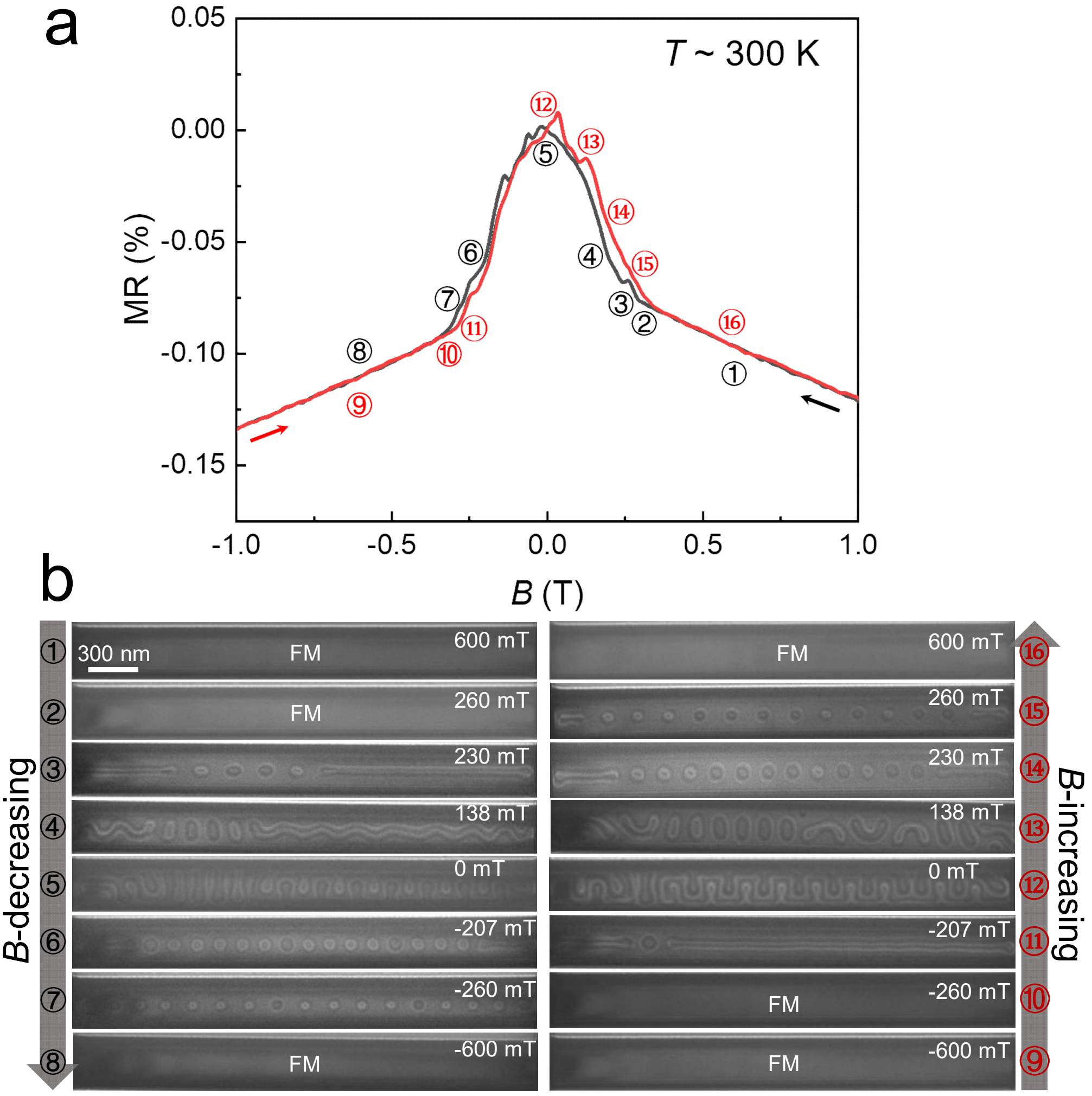

**Figure S13.** (a) MR as a function of the magnetic field at 300 K. The magnetic field range sweeps from 1 T to −1 T (black line) and back to 1 T (red line). The arrow denotes the direction of the applied magnetic field (b) Lorentz Fresnel images depicting the domain structure states under the same conditions as those in panel (a), with the numbers indicating the correspondence.

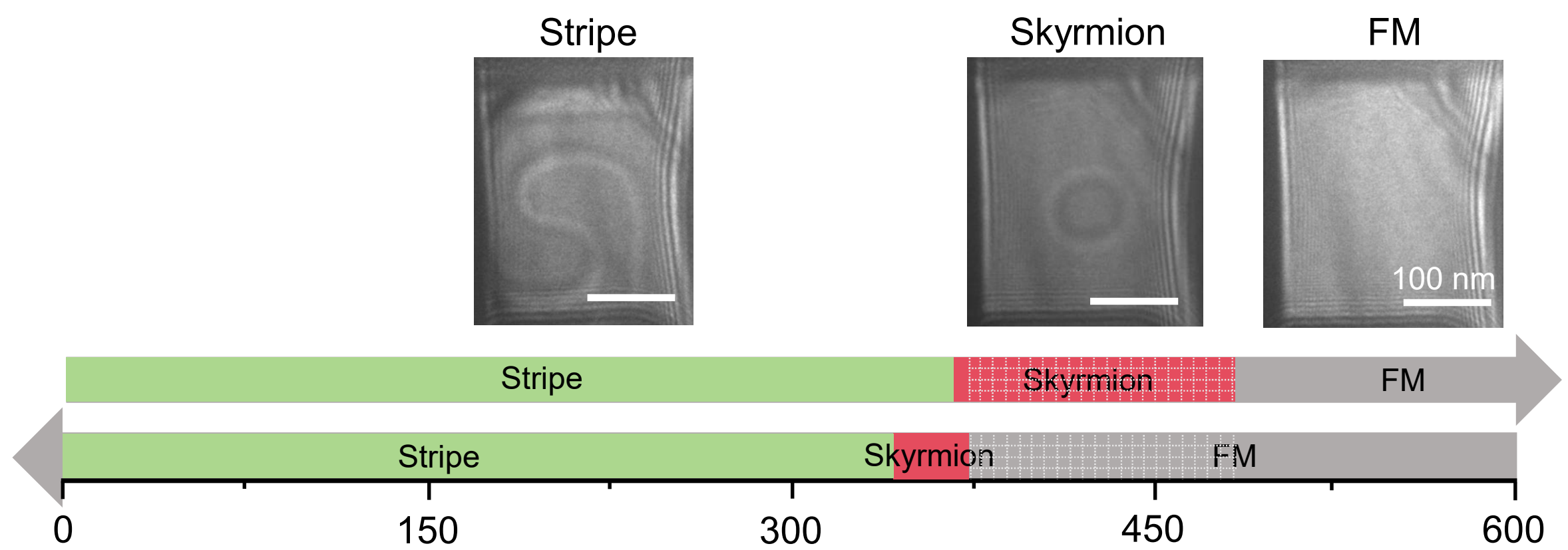

**Figure S14.** Magnetic phase diagram in increased and decreased field processes and representative defocused Fresnel contrasts of stripe, skyrmion, and ferromagnet (FM) in the $Mn_5Ge_{3+x}$ nanostructured unit. Skyrmion and FM are both stable phases in a field range of ~ 345-480 mT marked by a gray grid region.

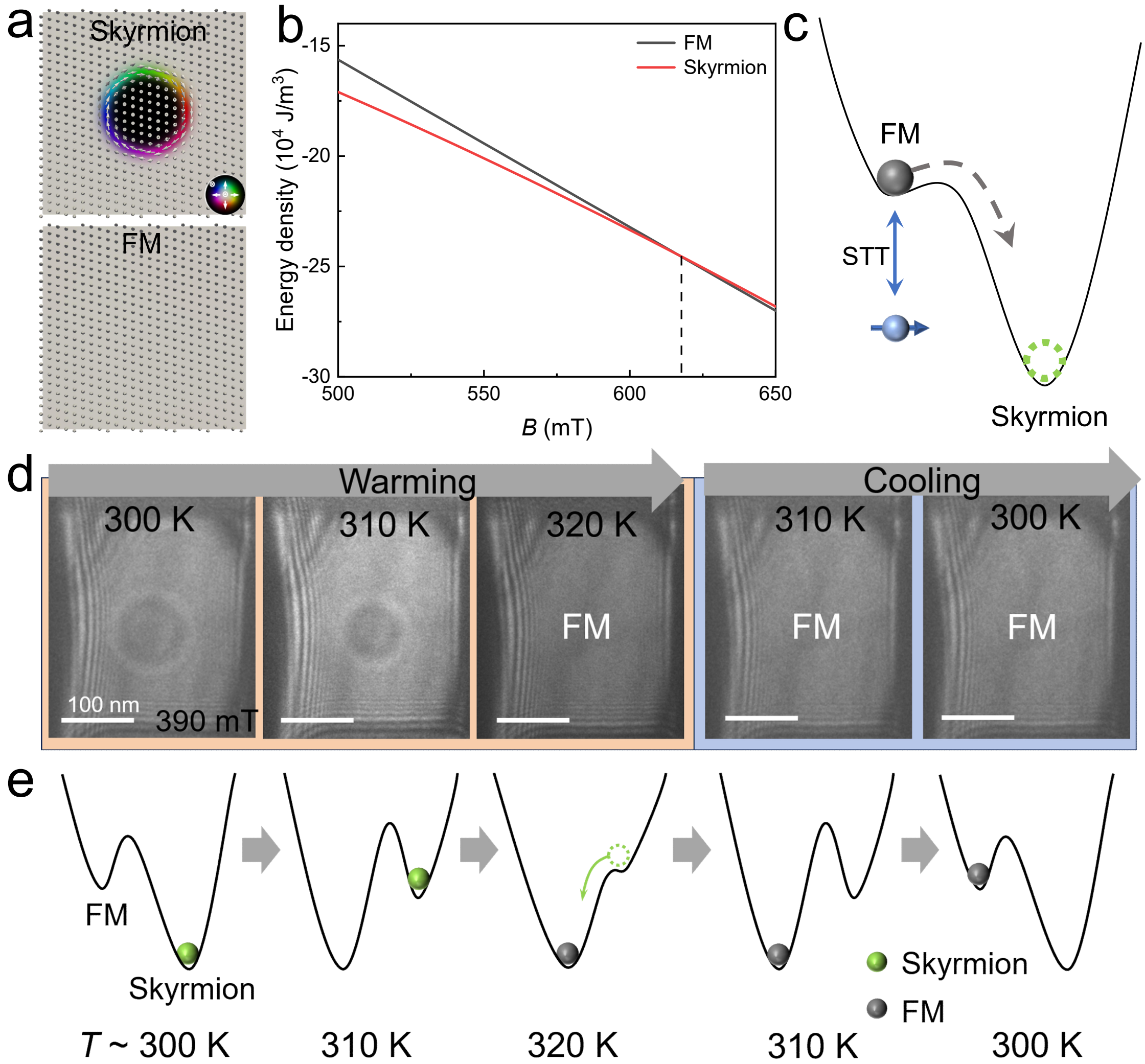

**Figure S15.** (a) Schematic configurations of the skyrmion, FM represented by green, and gray balls, respectively. (b) Simulated the energy density as a function of the magnetic field strength, when the initial states are set as skyrmion state and the ferromagnet (FM) state, respectively. (c) Schematic energetics during the FM-to-skyrmion magnetic transformation induced by the effect of spin-transfer torque (STT). (d) The magnetic evolution observed during the warming and cooling processes at a fixed magnetic field of 390 mT. (e) Schematic diagram of energy density profiles as the temperatures varies in a circular manner, corresponding to figure (d).

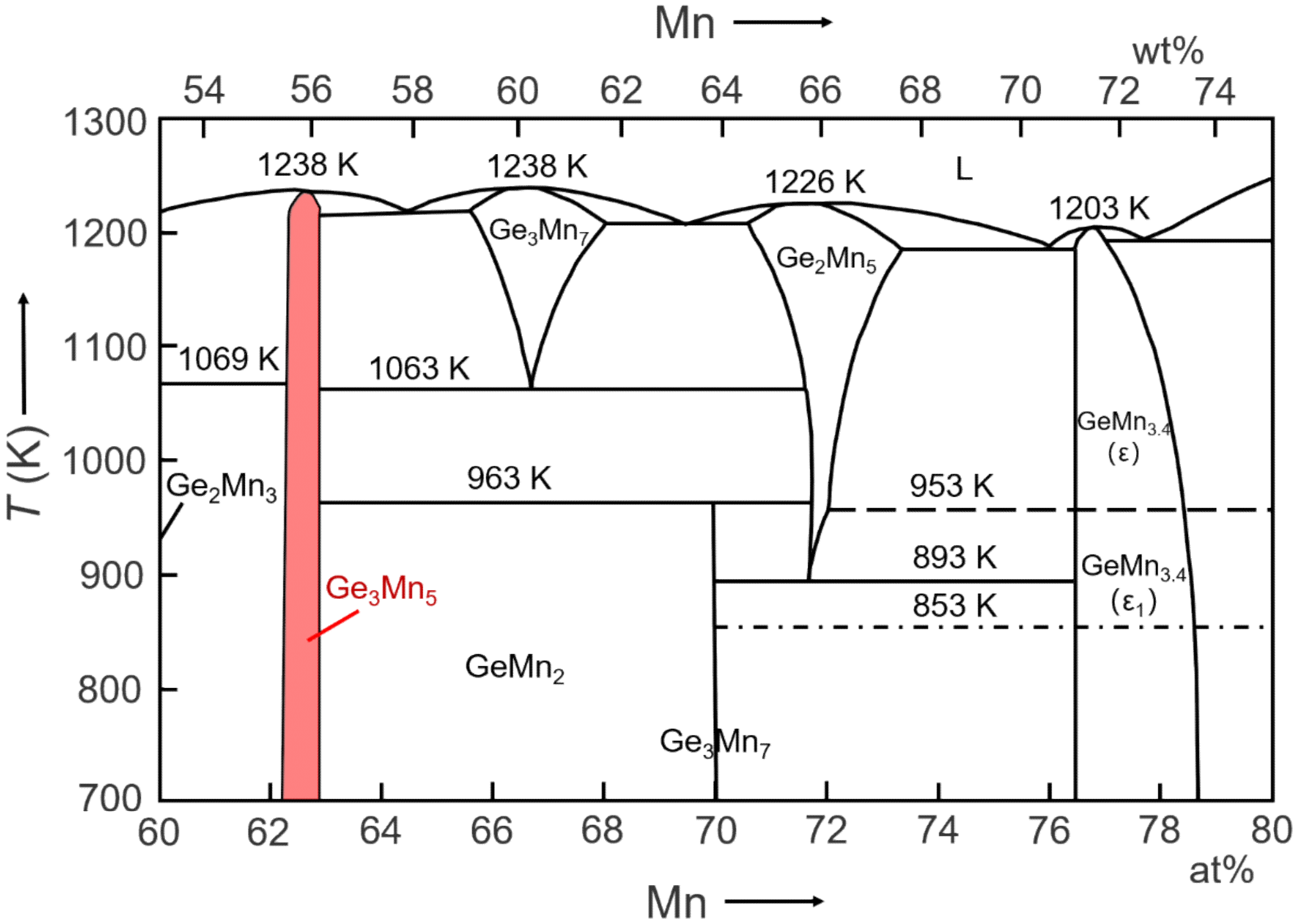

**Figure S16.** Partial phase diagram from 60% to 80% of at% Mn[54]. The red area indicates the $Mn_5Ge_3$ region.